\documentclass[sigconf]{acmart}

\usepackage{booktabs}
\usepackage{tabularx}
\usepackage{adjustbox}
\usepackage{multirow}
\usepackage{makecell}
\usepackage{xcolor}
\usepackage{colortbl}
\AtBeginDocument{%
  }

\setcopyright{acmlicensed}
\copyrightyear{2025}
\acmYear{2025}
\setcopyright{acmlicensed}\acmConference[CIKM '25]{Proceedings of the 34th ACM International Conference on Information and Knowledge Management}{November 10--14, 2025}{Seoul, Republic of Korea}
\acmBooktitle{Proceedings of the 34th ACM International Conference on Information and Knowledge Management (CIKM '25), November 10--14, 2025, Seoul, Republic of Korea}
\acmDOI{10.1145/3746252.3761170}
\acmISBN{979-8-4007-2040-6/2025/11}

\begin{document}

\title{UniECS: \underline{Uni}fied Multimodal \underline{E}-\underline{C}ommerce \underline{S}earch Framework with Gated Cross-modal Fusion}

\author{Zihan Liang}
\authornote{Equal Contribution.}
\affiliation{
  \institution{Kuaishou Technology}
  \city{Hangzhou}
  \state{Zhejiang}
  \country{China}
}
\email{liangzih@seas.upenn.edu}

\author{Yufei Ma}
\authornotemark[1]
\affiliation{
  \institution{Kuaishou Technology}
  \city{Hangzhou}
  \state{Zhejiang}
  \country{China}
}
\email{martinichi@mail.nwpu.edu.cn}

\author{ZhiPeng Qian}
\authornotemark[1]
\affiliation{%
  \institution{Kuaishou Technology}
  \city{Hangzhou}
  \state{Zhejiang}
  \country{China}
}
\email{qianzhipeng@stu.xmu.edu.cn}

\author{Huangyu Dai}
\affiliation{
  \institution{Kuaishou Technology}
  \city{Hangzhou}
  \state{Zhejiang}
  \country{China}
}
\email{11931034@zju.edu.cn}

\author{Zihan Wang}
\affiliation{
  \institution{Kuaishou Technology}
  \city{Hangzhou}
  \state{Zhejiang}
  \country{China}
}
\email{2101213096@stu.pku.edu.cn}

\author{Ben Chen}
\authornote{Corresponding author.}
\affiliation{
  \institution{Kuaishou Technology}
  \city{Hangzhou}
  \state{Zhejiang}
  \country{China}
}
\email{benchen4395@gmail.com}

\author{Chenyi Lei}
\affiliation{
  \institution{Kuaishou Technology}
  \city{Hangzhou}
  \state{Zhejiang}
  \country{China}
}
\email{leichy@mail.ustc.edu.cn}

\author{Yuqing Ding}
\affiliation{
  \institution{Kuaishou Technology}
    \city{Beijing}
  \country{China}
}
\email{kokia.ding@gmail.com}

\author{Han Li}
\affiliation{
  \institution{Kuaishou Technology}
  \city{Beijing}
  \country{China}
}
\email{lihan08@kuaishou.com}

\renewcommand{\shortauthors}{Liang et al.}

\begin{abstract}
The growth of e-commerce has created substantial demand for multimodal search systems that process diverse visual and textual inputs.
Current e-commerce multimodal retrieval systems face two key limitations: they optimize for specific tasks with fixed modality pairings, and lack comprehensive benchmarks for evaluating unified retrieval approaches.
To address these challenges, we introduce UniECS, a unified multimodal e-commerce search framework that handles all retrieval scenarios across image, text, and their combinations.
Our work makes three key contributions.
First, we propose a flexible architecture with a novel gated multimodal encoder that uses adaptive fusion mechanisms. 
This encoder integrates different modality representations while handling missing modalities.
Second, we develop a comprehensive training strategy to optimize learning. 
It combines cross-modal alignment loss (CMAL), cohesive local alignment loss (CLAL), intra-modal contrastive loss (IMCL), and adaptive loss weighting.
Third, we create M-BEER, a carefully curated multimodal benchmark containing 50K product pairs for e-commerce search evaluation.
Extensive experiments demonstrate that UniECS consistently outperforms existing methods across four e-commerce benchmarks with fine-tuning or zero-shot evaluation. 
On our M-BEER bench, UniECS achieves substantial improvements in cross-modal tasks (up to 28\% gain in R@10 for text-to-image retrieval) while maintaining parameter efficiency (0.2B parameters) compared to larger models like GME-Qwen2VL (2B) and MM-Embed (8B).
Furthermore, we deploy UniECS in the e-commerce search platform of Kuaishou Inc. across two search scenarios, achieving notable improvements in Click-Through Rate (+2.74\%) and Revenue (+8.33\%).
The comprehensive evaluation demonstrates the effectiveness of our approach in both experimental and real-world settings.
Corresponding codes, models and datasets will be made publicly available at \url{https://github.com/qzp2018/UniECS}.
\end{abstract}




\begin{teaserfigure}
\centering
\includegraphics[width=0.95\textwidth]{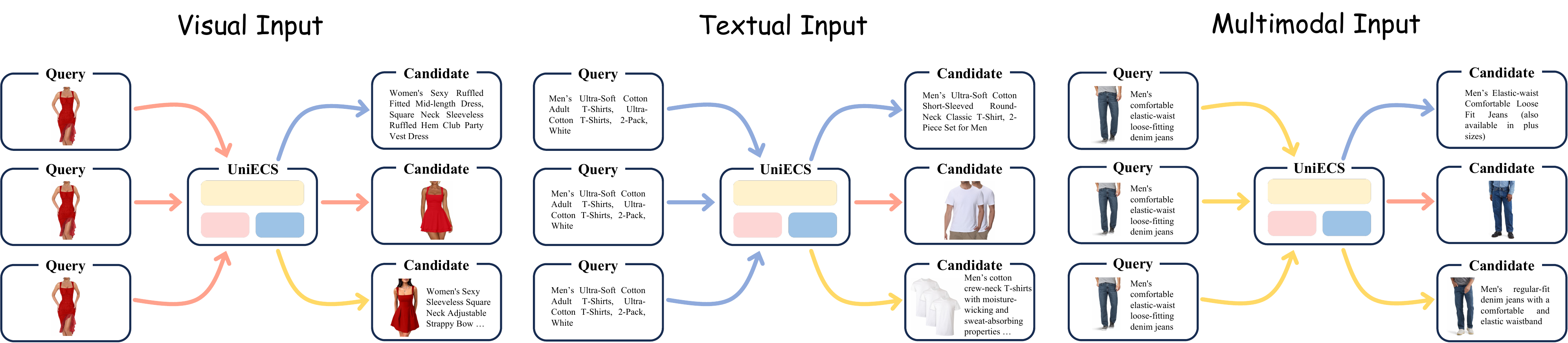}
\caption{Illustration of different retrieval settings in our universal multimodal retrieval task.}
\label{fig:teaser}
\end{teaserfigure}
\maketitle

\section{Introduction}
\par
E-commerce platforms have experienced substantial growth in recent years, creating demand for search systems that can process diverse user inputs. 
Traditional approaches in e-commerce typically operate within isolated modalities: text-to-text matching for keywords, image-to-image similarity for visual content, or simple multimodal-to-multimodal matching.
These systems lack cross-modal retrieval capabilities, searching with any modality type and retrieving results in any modality combination, which limits their effectiveness as shopping behaviors evolve~\cite{Zhu2024MIMAmazonKDD24}.  
Today's customers express their intent through textual descriptions, reference images, or combinations of both. For instance, a shopper might upload a photo of a jacket while specifying "in leather material" or "with fewer pockets" as refinements. 
Advancements in vision-language models (VLM)~\cite{li2023blip2,yang2023gpt4v,agrawal2024pixtral12b,dai2024nvlmopenfrontierclassmultimodal,wang2024qwen2vl,wu2024deepseekvl2,alayrac2022flamingo} and agents~\cite{yao2023react,hong2024metagpt,Zhao2024ExpelAgent,kim2024mdagents,Deng2023MIND2WEB,liang2024poir} have further elevated expectations for search capabilities. 
This shift in search behavior necessitates a flexible framework that can handle multiple input types throughout the shopping journey.

\par
Current multimodal retrieval systems in e-commerce face two key limitations. 
First, most frameworks are optimized for specific retrieval tasks with fixed query-candidate modality, lacking the flexibility to process arbitrary inputs. 
Even powerful models like CLIP~\cite{radford2021clip} and BLIP~\cite{li2022blip}, which excel at cross-modal alignment, require customized adaptations for different search scenarios. 
Second, the field lacks comprehensive benchmarks for evaluating multimodal search performance. This makes it difficult to assess unified retrieval approaches across various query-candidate combinations.

\par
Recent research has proposed several approaches to address unified retrieval challenges. 
UniIR~\cite{Wei2024uniir} introduces an instruction-following framework that handles varied datasets but lacks domain-specific optimizations for e-commerce. 
Both MM-Embed~\cite{lin2025mmembed} and GME~\cite{zhang2024gme}, while effective for general retrieval tasks, present two limitations: 
they are trained on general-domain data without specialization for product search requirements, and 
their large parameter counts (8B and 2B respectively) create computational demands that hinder deployment in latency-sensitive environments.

\par
To address these challenges, we propose UniECS, a \textbf{Uni}fied Multimodal \textbf{E}-\textbf{C}ommerce \textbf{S}earch framework that handles all modality combinations. 
Our approach introduces a flexible architecture with three key innovations: First, we design a gated multimodal encoder that employs adaptive fusion mechanisms to integrate visual and textual representations while gracefully handling missing modality scenarios. This encoder enables seamless transitions between single-modal and multimodal operations without architectural modifications. Second, we develop a comprehensive training strategy that incorporates CMAL, CLAL, and IMCL to optimize representation learning across different modalities. Third, we implement an adaptive loss weighting scheme that dynamically balances the contribution of each loss component. Additionally, to address the lack of comprehensive benchmarks in this field, we create M-BEER, a carefully curated multimodal benchmark containing 50K product pairs specifically for e-commerce search evaluation across nine distinct retrieval scenarios.

\par
Experimental evaluation demonstrates that UniECS consistently outperforms existing methods across multiple benchmarks while maintaining computational efficiency with only 0.2B parameters compared to GME-Qwen2VL's 2B and MM-Embed's 8B. 
On our newly introduced M-BEER benchmark, UniECS achieves substantial improvements in cross-modal tasks, with up to 28\% higher R@10 for text-to-image retrieval compared to the best baseline. 
The model also maintains competitive performance in single-modal scenarios, demonstrating its versatility across different search patterns.
Beyond experimental evaluation, we deploy UniECS in Kuaishou Inc.'s e-commerce platform across image search and same-style item retrieval scenarios. 
The deployment yields notable business improvements including 2.74\% improvement in Click-Through Rate, and 8.33\% growth Rrevenue, validating the practical effectiveness of our approach.
Corresponding codes, models and datasets will be made publicly available for community study.

\par
Our contributions are summarized as follows:
\begin{itemize}
    \item \textbf{Unified Architecture:} We develop a flexible framework that processes diverse visual and textual inputs through a single pipeline, supporting arbitrary modality combinations for both queries and candidates in e-commerce search.
    
    \item \textbf{Technical Advances:} We introduce a gated multimodal encoder with adaptive fusion capabilities that handles missing modalities effectively, coupled with a training strategy that enhances cross-modal alignment and representation quality through specialized loss functions.
    
    \item \textbf{Benchmark:} We create M-BEER, a comprehensive benchmark containing 50K product pairs for evaluating e-commerce search. Each sample contains trigger text, trigger image, recall text, and recall image, enabling standardized evaluation of nine distinct retrieval scenarios illustrated in Figure~\ref{fig:teaser}.
\end{itemize}

\begin{figure*}[th]
  \includegraphics[width=0.85\textwidth]{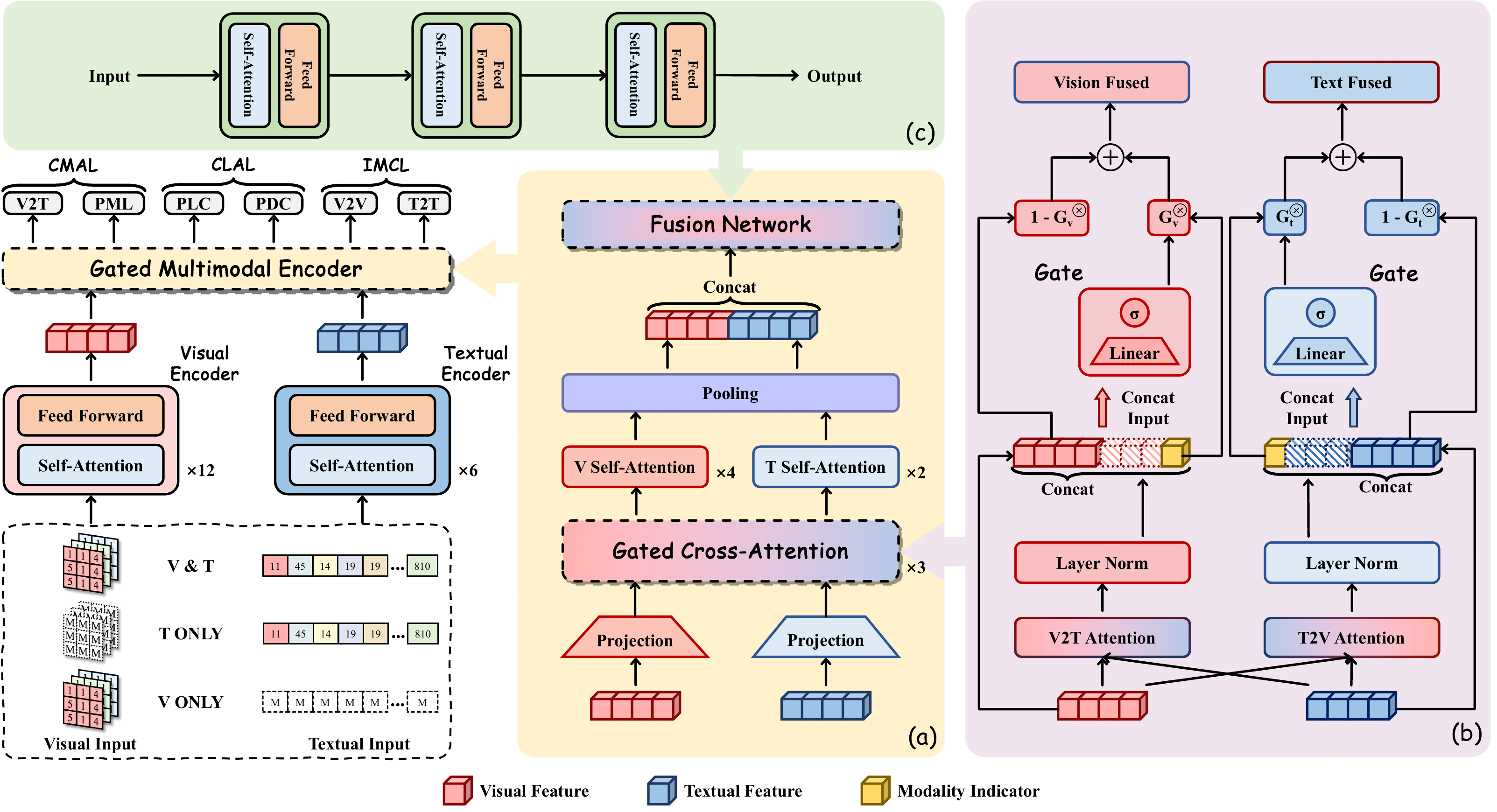}
  \caption{The illustration of our proposed UniECS method, focusing on arbitrary retrieval for visual and textual modalities. The core innovations of UniECS consist of a gated multimodal encoder (a), a gated cross-attention layer (b), and a fusion network (c).}
  \Description{Enjoying the baseball game from the third-base
  seats. Ichiro Suzuki preparing to bat.}
  \label{fig:method}
\end{figure*}

\section{Related Work}

\subsection{Multimodal Retrieval}
\par Multimodal retrieval has evolved from isolated modality approaches to integrated frameworks capable of handling diverse inputs. 
Early research focused on single-modal or cross-modal retrieval through dual-encoder architectures such as ALIGN~\cite{jia2021align}, CLIP~\cite{radford2021clip}, ALBEF~\cite{li2021albef} and BLIP~\cite{li2022blip}. 
These models established effective alignment between visual and textual representations but were typically optimized for specific tasks with fixed query-candidate modality pairings, limiting their flexibility in e-commerce applications.

\par 
In e-commerce, the challenge of effective product retrieval has driven significant innovations. 
Pure image-to-image matching often captures irrelevant background features, resulting in imprecise search results. 
Recent approaches address these limitations through two primary strategies: (1) feature fusion mechanisms based on more flexible architectures as seen in UniVL-DR~\cite{liu2023univldr} and instruction-following frameworks like UniIR that can handle various modality combinations for search, and (2) visual plugin modules integrated with text embedding systems like MARVEL~\cite{zhou-etal-2024-marvel} and VISTA~\cite{zhou-etal-2024-vista}. However, many of these solutions still lack domain-specific optimizations for e-commerce or require complex adaptations for different search scenarios.

\par 
With the emergence of Multimodal Large Language Models (MLLMs), the field has expanded toward Universal Multimodal Retrieval (UMR). 
Models such as E5-V~\cite{jiang2024e5v} and GME demonstrate that fine-tuning MLLMs with diverse multimodal data creates effective universal embeddings. 
GME showed that balancing single-modal, cross-modal, and fused-modal training data improves performance across various retrieval scenarios. 
MM-Embed~\cite{lin2025mmembed} similarly applies large-scale models to achieve flexible representation learning. 
However, these approaches present challenges in applications: they typically require huge computational resources and lack domain-specific optimization for product search requirements.

\par 
Our UniECS framework addresses these limitations by integrating visual and textual information through an adaptive gating mechanism that enables efficient cross-modal matching while maintaining a parameter-efficient architecture (0.2B parameters). 
This approach allows customers to refine product searches using both visual references and textual specifications, resulting in more accurate and flexible shopping experiences.

\subsection{Deep Metric Learning}
\par 
Deep metric learning forms the foundation of modern similarity-based retrieval systems. 
It creates embedding spaces where semantically similar objects cluster together while dissimilar objects remain separated.
Researchers have developed various loss functions to train embedding models, generally categorized into: pair-based approaches like contrastive~\cite{Raia2006contrastiveloss} and triplet losses~\cite{Florian2015tripletloss}, classification-based losses~\cite{Mao2023CELoss}, and proxy-anchor mechanisms~\cite{Kim2020ProxyAnchorLoss}.

\par 
Our contribution extends traditional metric learning with specialized objective functions designed for multimodal e-commerce retrieval. 
CMAL ensures semantic consistency between different modality representations, while CLAL maintains representation structure consistency across modalities. Additionally, IMCL preserves strong single-modal performance. This comprehensive loss design, combined with our adaptive gating mechanism, enables UniECS to efficiently handle diverse retrieval scenarios.

\section{Method}
\par
In this section, we present our framework, which supports flexible combinations of visual, textual, and multimodal inputs. 
We first introduce the overall architecture, followed by the feature extraction and fusion components. 
Then, we describe our adaptive gating mechanism for handling missing modalities. 
Finally, we detail our comprehensive loss design that enhances cross-modal alignment, feature quality, and single-modality retrieval performance.

\subsection{Framework Overview}
\par
Our framework addresses the complex demands of e-commerce search, where users may provide queries in visual form (product images), textual form (product descriptions), or both. As shown in Figure~\ref{fig:method}, the framework consists of three main components: 
(1) single-modal encoders for visual and textual inputs, 
(2) a multimodal fusion encoder with an adaptive gating mechanism, 
and (3) a comprehensive training objective with multiple specialized loss functions.
The input can be any combination of image and text, making it flexible for various scenarios. 

\subsection{Feature Extraction}
\par
For visual inputs, we employ a Vision Transformer (ViT) as the backbone, which processes input images of size $224 \times 224$ pixels.
The visual encoder outputs token-level features, denoted as $\mathbf{V} \in \mathbb{R}^{N_v \times D_v}$, where $N_v$ is the number of visual tokens and $D_v$ is the feature dimension.
These features capture visual characteristics of products, including shapes, colors, and patterns.
\par
For textual inputs, we adopt a BERT-based encoder that processes tokenized product descriptions or queries.
The textual encoder outputs token-level features, denoted as $\mathbf{T} \in \mathbb{R}^{N_t \times D_t}$, where $N_t$ is the number of text tokens and $D_t$ is the feature dimension.

\subsection{Feature Fusion}
\par
We propose a multimodal encoder with an adaptive gating mechanism for feature fusion. This encoder integrates visual and textual features through specialized components, as illustrated in Figure~\ref{fig:method} (a) and (b). The encoder consists of three key components: cross-modal attention, adaptive gating, and self-attention layers.

\par
First, we project features from both modalities to a common embedding space:
\begin{equation}\small
\mathbf{V}' = \text{Proj}_v(\mathbf{V}), \quad \mathbf{T}' = \text{Proj}_t(\mathbf{T}),
\end{equation}
where $\text{Proj}_v$ and $\text{Proj}_t$ are projection layers for visual and textual features, respectively.

\par
Next, we apply cross-modal attention to enable information exchange between modalities. For brevity, we present the visual modality operations below, with textual operations following the same pattern:
\begin{equation}\small
\mathbf{V}_{\text{attn}} = \text{LayerNorm}(\text{V2T-Attention}(\mathbf{V}', \mathbf{T}')),
\end{equation}
where the attention mechanism allows the visual modality to focus on relevant textual information. The first argument serves as query while the second provides keys and values.

\par
The adaptive gating mechanism determines how much cross-attention information to incorporate, enabling the model to handle missing modalities and balance contributions from each modality:
\begin{equation}\small
\mathbf{G}_v = \sigma(\mathbf{W}_{v}([\mathbf{V}';\mathbf{V}_{\text{attn}};\mathbf{E}_v]) + \mathbf{b}_v),
\end{equation}
where $\sigma$ is the sigmoid function, $[\cdot;\cdot;\cdot]$ represents feature concatenation, and $\mathbf{E}_v$ is a modality indicator signaling the presence or absence of the visual modality.

\par
The final gated output adaptively combines original features with cross-attended features:
\begin{equation}\small
\mathbf{V}_{\text{g}} = \mathbf{G}_v \odot \mathbf{V}_{\text{attn}} + (1 - \mathbf{G}_v) \odot \mathbf{V}',
\end{equation}
where $\odot$ denotes element-wise multiplication. This weighting mechanism handles cases where one modality is missing or noisy, enabling smooth transitions between single-modal and multimodal operations without architectural changes.

\par
Finally, the outputs are processed through modality-specific self-attention blocks to capture intra-modal dependencies:
\begin{equation}\small
\mathbf{V}'' = \text{V-Self-Attention}(\mathbf{V}_{\text{g}}).
\end{equation}
The textual modality follows identical operations with corresponding parameters and notations.

\subsection{Global Feature Generation}
\par
After the gated multimodal encoding stage, we generate normalized global representations through our fusion network, as illustrated in Figure~\ref{fig:method} (c). 
This network processes the pooled features from the gated encoder through three Transformer blocks:
\begin{equation}\small
\mathbf{f} = \text{FusionNetwork}([\text{Pool}(\mathbf{V}''); \text{Pool}(\mathbf{T}'')]).
\end{equation}

\par
Our framework adapts to different input configurations:
\begin{itemize}
\item When both modalities are present, it can generate three representations: visual-only ($\mathbf{v}$), text-only ($\mathbf{t}$), and multimodal ($\mathbf{f}$). Single-modal representations are obtained by setting the other modality's features and indicators to zero.
\item When only one modality is available, it automatically sets missing modality features and indicators to zero, with the gating mechanism adapting to enable seamless single-modal operation.
\end{itemize}
\par
This approach enables a unified pipeline that handles any modality combination without architectural changes. All representations are L2-normalized before being used for retrieval.

\subsection{Loss Design}
Our framework utilizes a comprehensive training objective that addresses three key aspects of multimodal retrieval: cross-modal alignment, feature structure consistency, and intra-modal discrimination. The total loss function is formulated as:
\begin{equation}\small
\mathcal{L} = \mathcal{L}_{\text{CMAL}} + \mathcal{L}_{\text{CLAL}} + \mathcal{L}_{\text{IMCL}},
\end{equation}
where $\mathcal{L}_{\text{CMAL}}$ represents cross-modal alignment losses, $\mathcal{L}_{\text{CLAL}}$ denotes multimodal consistency loss, and $\mathcal{L}_{\text{IMCL}}$ refers to intra-modal contrastive loss.

\subsubsection{Cross-Modal Alignment Loss (CMAL)}
CMAL ensure semantic consistency between different modality representations, enabling effective cross-modal retrieval. 
This group includes:
\paragraph{Visual-Textual Contrastive Loss (V2T)} This loss aligns visual and textual representations:
\begin{equation}\small
\mathcal{L}_{\text{V2T}} = -\frac{1}{N}\sum_{i=1}^{N} \log \frac{\exp(\mathbf{v}_i \cdot \mathbf{t}_i / \tau)}{\sum_{j=1}^{N} \exp(\mathbf{v}_i \cdot \mathbf{t}_j / \tau)},
\end{equation}
where $\mathbf{v}_i$ and $\mathbf{t}_i$ are paired visual and textual representations, and $\tau$ is a temperature parameter.
\paragraph{Product Matching Loss (PML)} This loss maximizes similarity between different representations (multimodal, visual, textual) of paired similar products:
\begin{equation}\small
\begin{aligned}
\mathcal{L}_{\text{PML}} = & -\frac{1}{N}\sum_{i=1}^{N} [\log \frac{\exp(\mathbf{f}_i^{(1)} \cdot \mathbf{f}_i^{(2)} / \tau)}{\sum_{j=1}^{N} \exp(\mathbf{f}_i^{(1)} \cdot \mathbf{f}_j^{(2)} / \tau)} + \\
&\log \frac{\exp(\mathbf{v}_i^{(1)} \cdot \mathbf{f}_i^{(2)} / \tau)}{\sum_{j=1}^{N} \exp(\mathbf{v}_i^{(1)} \cdot \mathbf{f}_j^{(2)} / \tau)} + 
\log \frac{\exp(\mathbf{t}_i^{(1)} \cdot \mathbf{f}_i^{(2)} / \tau)}{\sum_{j=1}^{N} \exp(\mathbf{t}_i^{(1)} \cdot \mathbf{f}_j^{(2)} / \tau)}],
\end{aligned}
\end{equation}
where $\mathbf{f}_i^{(1)}$ and $\mathbf{f}_i^{(2)}$ are multimodal representations from two similar products.
The total cross-modal alignment loss is:
\begin{equation}\small
\mathcal{L}_{\text{CMAL}} = \lambda_{\text{V2T}}\mathcal{L}_{\text{V2T}} + \lambda_{\text{PML}}\mathcal{L}_{\text{PML}},
\end{equation}
where $\lambda_{\text{V2T}}$ and $\lambda_{\text{PML}}$ are weighting parameters.

\subsubsection{Cohesive Local Alignment Loss (CLAL)}
CLAL improves how well features can distinguish between different items (Locality) while ensuring similar items stay grouped together (Cohesion). 
This group includes:
\paragraph{Product Distinctiveness Consistency Loss (PDC)}
This loss ensures three modal representations for a same item maintain smaller distance:

\begin{equation}\small
\begin{aligned}
\mathcal{L}_{\text{PDC}} = \frac{1}{N}\sum_{i=1}^{N} \sum_{j=1}^{N} & [\max(0, \alpha_2 + s_{ij}^{M2M} - s_{ii}^{V2M}) + \\
& \max(0, \alpha_2 + s_{ij}^{M2M} - s_{ii}^{T2M})],
\end{aligned}
\end{equation}

where $s_{ij}^{M2M}$ is the similarity between multimodal representations, $s_{ii}^{V2M}$ and $s_{ii}^{T2M}$ are the similarities between visual/text and multimodal representations, and $\alpha_2$ is a margin parameter.
\paragraph{Product Locality Consistency Loss (PLC)}
This loss reduces interference of irrelevant attributes across different embeddings:
\begin{equation}\small
\begin{aligned}
\mathcal{L}_{\text{PLC}} = \frac{1}{\mathcal{N}_K(i)}\sum_{i=1}^{N} \sum_{j \in \mathcal{N}_{K}(i)} [& \max(0, -\alpha_3 + (s_{ij}^{V2M} - s_{ij}^{M2M})^2) + \\
& \max(0, -\alpha_3 + (s_{ij}^{T2M} - s_{ij}^{M2M})^2) + \\
& \max(0, -\alpha_3 + (s_{ij}^{V2M} - s_{ij}^{T2M})^2)],
\end{aligned}
\end{equation}
where $\mathcal{N}_K(i)$ represents the top $K$ most similar candidates for sample $i$, and $\alpha_3$ is a consistency threshold.
The total cohesive local alignment loss is:
\begin{equation}\small
\mathcal{L}_{\text{CLAL}} = \lambda_{\text{PDC}}\mathcal{L}_{\text{PDC}} + \lambda_{\text{PLC}}\mathcal{L}_{\text{PLC}},
\end{equation}
where $\lambda_{\text{PDC}}$ and $\lambda_{\text{PLC}}$ are weighting parameters.
\subsubsection{Intra-Modal Contrastive Loss (IMCL)}
IMCL enhance the discriminative power of single-modal representations, ensuring high-quality results even in single-modality retrieval scenarios. This group includes:
\paragraph{Visual-to-Visual Contrastive Loss (V2V)} This loss combines standard contrastive learning with hard negative mining:
\begin{equation}\small
\mathcal{L}_{\text{V2V}} = \mathcal{L}_{\text{std}} + \mathcal{L}_{\text{hard}},
\end{equation}
where:
\begin{equation}\small
\mathcal{L}_{\text{std}} = -\frac{1}{N}\sum_{i=1}^{N} \log \frac{\exp(\mathbf{v}_i^{(1)} \cdot \mathbf{v}_i^{(2)} / \tau_v)}{\sum_{j=1}^{N} \exp(\mathbf{v}_i^{(1)} \cdot \mathbf{v}_j^{(2)} / \tau_v)},
\end{equation}
\begin{equation}\small
\mathcal{L}_{\text{hard}} = \frac{1}{\mathcal{N}_K(i)}\sum_{i=1}^{N} \sum_{j \in \mathcal{H}_i} \max(0, \alpha_4 + s_{ij}^{V2V} - s_{ii}^{V2V}),
\end{equation}
where $\tau_v$ is the temperature parameter, $\mathcal{H}_i$ is the set of hard negative candidates for sample $i$, and $\alpha_4$ is a margin parameter.
\paragraph{Text-to-Text Contrastive Loss (T2T)} This loss optimizes text representations:
\begin{equation}\small
\mathcal{L}_{\text{T2T}} = -\frac{1}{N}\sum_{i=1}^{N} \log \frac{\exp(\mathbf{t}_i^{(1)} \cdot \mathbf{t}_i^{(2)} / \tau_t)}{\sum_{j=1}^{N} \exp(\mathbf{t}_i^{(1)} \cdot \mathbf{t}_j^{(2)} / \tau_t)},
\end{equation}
where $\tau_t$ is the temperature parameter, typically set lower to enhance precision in text matching. 
$\mathbf{t}_i^{(1)}$ and $\mathbf{t}_i^{(2)}$ are the titles for product pair $i$.
The total intra-modal contrastive loss is:
\begin{equation}\small
\mathcal{L}_{\text{IMCL}} = \lambda_{\text{V2V}}\mathcal{L}_{\text{V2V}} + \lambda_{\text{T2T}}\mathcal{L}_{\text{T2T}},
\end{equation}
where $\lambda_{\text{V2V}}$ and $\lambda_{\text{T2T}}$ are weighting parameters.

\subsection{Adaptive Loss Weighting}
To balance the contributions of different loss components, we implement an adaptive weighting scheme that dynamically adjusts the weight of each loss based on its gradient magnitude and training progress. 
This technique ensures that all aspects of the model are optimized effectively, preventing any single loss from dominating the training process.
The adaptive weights are updated at each training step using:

\begin{equation}\small
\lambda_i^{(t+1)} = \beta \cdot \frac{\lambda_i^{(t)} \cdot \exp\left(g_i\right)}{\sum_k \lambda_k^{(t)} \cdot \exp\left(g_k\right)} + (1-\beta) \cdot \lambda_i^{(t)},
\end{equation}

where $g_i = \frac{|\nabla_\theta \mathcal{L}_i|}{\sum_j |\nabla_\theta \mathcal{L}_j|}$, $\lambda_i^{(t)}$ is the weight for the $i$-th loss component at step $t$, $\beta$ is a balancing hyper-parameter, and $|\nabla_\theta \mathcal{L}_i|$ is the gradient magnitude of the $i$-th loss.
This adaptive weighting mechanism enables our framework to automatically focus on the most challenging aspects throughout the training process.

\section{Experiments}

\begin{table*}[ht]
\centering
\scriptsize
\renewcommand{\arraystretch}{0.85} 
\caption{Benchmarking multimodal retrieval performance on M-BEER Bench and eSSPR Bench. we report results by Recall@1, Recall@5, and Recall@10 metrics. \textbf{Bold}: top-1 performance. \underline{Underline}: top-2.}
\begin{adjustbox}{max width=0.85\textwidth}
\begin{tabular*}{0.9\textwidth}{@{\extracolsep{\fill}}ccccc|cccccc@{\extracolsep{\fill}}}
    \toprule
    \multirow{2}{*}[0ex]{\textbf{Task}} & \multirow{2}{*}[0ex]{\textbf{Method}} & \multirow{2}{*}[0ex]{\textbf{Venue}} & \multirow{2}{*}[0ex]{\textbf{Emb. Dim.}} & \multirow{2}{*}[0ex]{\textbf{Size}} & \multicolumn{3}{c}{\textbf{M-BEER Bench}} & \multicolumn{3}{c}{\textbf{eSSPR Bench (zero-shot)}} \\
    \cmidrule(lr){6-8}
    \cmidrule(lr){9-11}
    ~ & ~ & ~ & ~ & ~ & \textbf{R@1} & \textbf{R@5} & \textbf{R@10} & \textbf{R@1} & \textbf{R@5} & \textbf{R@10} \\
    \midrule
    \multirow{8}{*}{$ q_t \to c_v$} & $\text{CLIP}_{SF}$ & ECCV2024 & 768 & 0.4B &  0.02 & 0.07 & 0.15 & 0.05 & 0.18 & 0.28 \\
    ~ & $\text{CLIP}_{FF}$ & ECCV2024 & 768 & 0.4B &  0.03 & 0.09 & 0.17 & 0.08 & 0.25 & 0.35 \\
    ~ & $\text{BLIP}_{SF}$ & ECCV2024 & 768 & 0.9B &  0.05 & 0.12 & 0.24 & 0.10 & 0.32 & 0.42 \\
    ~ & $\text{BLIP}_{FF}$ & ECCV2024 & 768 & 0.9B &  0.08 & 0.16 & 0.26 & 0.12 & 0.38 & 0.48 \\
    ~ & BGE-VL-large & - & 768 & 0.4B & 0.10 & 0.22 & 0.30 & 0.15 & 0.45 & 0.57 \\
    ~ & GME-Qwen2VL & CVPR2025 & 1536 & 2B & \underline{0.26} & \underline{0.48} & \underline{0.57} & \textbf{0.23} & \textbf{0.64} & \underline{0.76} \\
    ~ & MM-Embed & ICLR2025 & 4096 & 8B & 0.18 & 0.38 & 0.48 & 0.18 & 0.49 & 0.62 \\
    \cmidrule(lr){2-11}
    ~ & UniECS (Ours) & CIKM2025 & 256 & 0.2B & \textbf{0.36} & \textbf{0.74} & \textbf{0.85} & \underline{0.22} & \underline{0.63} & \textbf{0.77} \\
    \midrule
    \multirow{8}{*}{$q_t \to c_t$} & $\text{CLIP}_{SF}$ & ECCV2024 & 768 & 0.4B & 0.26 & 0.45 & 0.55 & 0.12 & 0.35 & 0.48 \\
    ~ & $\text{CLIP}_{FF}$ & ECCV2024 & 768 & 0.4B &  0.40 & 0.62 & 0.72 & 0.18 & 0.58 & 0.68 \\
    ~ & $\text{BLIP}_{SF}$ & ECCV2024 & 768 & 0.9B &  0.30 & 0.45 & 0.53 & 0.20 & 0.60 & 0.72 \\
    ~ & $\text{BLIP}_{FF}$ & ECCV2024 & 768 & 0.9B &  0.32 & 0.48 & 0.56 & 0.22 & 0.65 & 0.76 \\
    ~ & BGE-VL-large & - & 768 & 0.4B & 0.28 & 0.43 & 0.50 & 0.25 & 0.67 & 0.74 \\
    ~ & GME-Qwen2VL & CVPR2025 & 1536 & 2B & \underline{0.61} & \underline{0.90} & \underline{0.95} & \underline{0.42} & \underline{0.84} & \underline{0.92} \\
    ~ & MM-Embed & ICLR2025 & 4096 & 8B & 0.53 & 0.80 & 0.87 & 0.39 & 0.81 & 0.89 \\
    \cmidrule(lr){2-11}
    ~ & UniECS (Ours) & CIKM2025 & 256 & 0.2B & \textbf{0.62} & \textbf{0.93} & \textbf{0.98} & \textbf{0.44} & \textbf{0.94} & \textbf{0.98} \\ 
    \midrule
    \multirow{8}{*}{$q_t \to (c_v,c_t)$} & $\text{CLIP}_{SF}$ & ECCV2024 & 768 & 0.4B & 0.25 & 0.44 & 0.56 & 0.15 & 0.40 & 0.52 \\
    ~ & $\text{CLIP}_{FF}$ & ECCV2024 & 768 & 0.4B &  0.33 & 0.58 & 0.68 & 0.18 & 0.45 & 0.58 \\
    ~ & $\text{BLIP}_{SF}$ & ECCV2024 & 768 & 0.9B &  0.29 & 0.46 & 0.59 & 0.25 & 0.73 & 0.83 \\
    ~ & $\text{BLIP}_{FF}$ & ECCV2024 & 768 & 0.9B &  0.17 & 0.35 & 0.45 & 0.24 & 0.72 & 0.83 \\
    ~ & BGE-VL-large & - & 768 & 0.4B & 0.25 & 0.38 & 0.45 & 0.22 & 0.69 & 0.76 \\
    ~ & GME-Qwen2VL & CVPR2025 & 1536 & 2B & \underline{0.60} & \underline{0.89} & \underline{0.95} & 0.32 & \textbf{0.87} & \textbf{0.93} \\
    ~ & MM-Embed & ICLR2025 & 4096 & 8B & 0.54 & 0.82 & 0.89 & \underline{0.34} & 0.84 & 0.91 \\
    \cmidrule(lr){2-11}
    ~ & UniECS (Ours) & CIKM2025 & 256 & 0.2B & \textbf{0.63} & \textbf{0.94} & \textbf{0.98} & \textbf{0.37} & \underline{0.86} & \underline{0.92} \\ 
    \midrule
    \multirow{8}{*}{$q_v \to c_t$} & $\text{CLIP}_{SF}$ & ECCV2024 & 768 & 0.4B &  0.09 & 0.14 & 0.23 & 0.08 & 0.22 & 0.30 \\
    ~ & $\text{CLIP}_{FF}$ & ECCV2024 & 768 & 0.4B &  0.08 & 0.17 & 0.27 & 0.10 & 0.28 & 0.35 \\
    ~ & $\text{BLIP}_{SF}$ & ECCV2024 & 768 & 0.9B &  0.09 & 0.17 & 0.32 & 0.12 & 0.32 & 0.40 \\
    ~ & $\text{BLIP}_{FF}$ & ECCV2024 & 768 & 0.9B &  0.12 & 0.24 & 0.36 & 0.15 & 0.36 & 0.45 \\
    ~ & BGE-VL-large & - & 768 & 0.4B & 0.15 & 0.28 & 0.38 & 0.18 & 0.42 & 0.52 \\
    ~ & GME-Qwen2VL & CVPR2025 & 1536 & 2B & \underline{0.25} & \underline{0.50} & \underline{0.60} & \underline{0.26} & \textbf{0.66} & \textbf{0.78} \\
    ~ & MM-Embed & ICLR2025 & 4096 & 8B & 0.22 & 0.45 & 0.56 & 0.17 & 0.47 & 0.59 \\
    \cmidrule(lr){2-11}
    ~ & UniECS (Ours) & CIKM2025 & 256 & 0.2B & \textbf{0.38} & \textbf{0.77} & \textbf{0.87} & \textbf{0.27} & \underline{0.63} & \underline{0.77} \\ 
    \midrule
    \multirow{8}{*}{$q_v \to c_v$} & $\text{CLIP}_{SF}$ & ECCV2024 & 768 & 0.4B & 0.35 & 0.51 & 0.59 & 0.17 & 0.86 & 0.90 \\
    ~ & $\text{CLIP}_{FF}$ & ECCV2024 & 768 & 0.4B &  0.51 & 0.75 & 0.82 & 0.15 & 0.94 & 0.97 \\
    ~ & $\text{BLIP}_{SF}$ & ECCV2024 & 768 & 0.9B &  \textbf{0.65} & \underline{0.88} & \underline{0.94} & 0.15 & 0.95 & 0.97 \\
    ~ & $\text{BLIP}_{FF}$ & ECCV2024 & 768 & 0.9B &  0.62 & 0.86 & 0.92 & 0.20 & 0.96 & 0.98 \\
    ~ & BGE-VL-large & - & 768 & 0.4B & 0.23 & 0.43 & 0.54 & \textbf{0.34} & 0.91 & 0.96 \\
    ~ & GME-Qwen2VL & CVPR2025 & 1536 & 2B & 0.58 & 0.80 & 0.86 & 0.30 & \underline{0.98} & \underline{0.99} \\
    ~ & MM-Embed & ICLR2025 & 4096 & 8B & 0.55 & 0.78 & 0.85 & 0.28 & 0.97 & 0.98 \\
    \cmidrule(lr){2-11}
    ~ & UniECS (Ours) & CIKM2025 & 256 & 0.2B & \underline{0.63} & \textbf{0.89} & \textbf{0.95} & \underline{0.32} & \textbf{0.99} & \textbf{1.00} \\ 
    \midrule
    \multirow{8}{*}{$q_v \to (c_v,c_t)$} & $\text{CLIP}_{SF}$ & ECCV2024 & 768 & 0.4B & 0.28 & 0.44 & 0.53 & 0.21 & 0.81 & 0.86 \\
    ~ & $\text{CLIP}_{FF}$ & ECCV2024 & 768 & 0.4B &  0.30 & 0.56 & 0.68 & 0.26 & 0.90 & 0.94 \\
    ~ & $\text{BLIP}_{SF}$ & ECCV2024 & 768 & 0.9B &  0.61 & 0.85 & 0.92 & 0.26 & \underline{0.98} & \underline{0.99} \\
    ~ & $\text{BLIP}_{FF}$ & ECCV2024 & 768 & 0.9B &  \underline{0.62} & \underline{0.86} & \underline{0.93} & 0.29 & 0.96 & 0.98 \\
    ~ & BGE-VL-large & - & 768 & 0.4B & 0.27 & 0.52 & 0.63 & 0.25 & 0.78 & 0.88 \\
    ~ & GME-Qwen2VL & CVPR2025 & 1536 & 2B & 0.50 & 0.77 & 0.85 & \underline{0.33} & 0.95 & 0.98 \\
    ~ & MM-Embed & ICLR2025 & 4096 & 8B & 0.52 & 0.75 & 0.83 & 0.28 & 0.92 & 0.96 \\
    \cmidrule(lr){2-11}
    ~ & UniECS (Ours) & CIKM2025 & 256 & 0.2B & \textbf{0.64} & \textbf{0.92} & \textbf{0.97} & \textbf{0.35} & \textbf{0.99} & \textbf{1.00} \\ 
    \midrule
    \multirow{8}{*}{$(q_v,q_t) \to c_t$} & $\text{CLIP}_{SF}$ & ECCV2024 & 768 & 0.4B &  0.24 & 0.45 & 0.53 & 0.18 & 0.42 & 0.55 \\
    ~ & $\text{CLIP}_{FF}$ & ECCV2024 & 768 & 0.4B &  0.31 & 0.56 & 0.67 & 0.22 & 0.52 & 0.65 \\
    ~ & $\text{BLIP}_{SF}$ & ECCV2024 & 768 & 0.9B &  0.29 & 0.44 & 0.54 & 0.25 & 0.58 & 0.68 \\
    ~ & $\text{BLIP}_{FF}$ & ECCV2024 & 768 & 0.9B &  0.32 & 0.52 & 0.62 & 0.28 & 0.60 & 0.72 \\
    ~ & BGE-VL-large & - & 768 & 0.4B & 0.25 & 0.40 & 0.48 & 0.22 & 0.69 & 0.76 \\
    ~ & GME-Qwen2VL & CVPR2025 & 1536 & 2B & \underline{0.58} & \underline{0.90} & \underline{0.95} & \underline{0.35} & \underline{0.85} & \underline{0.94} \\
    ~ & MM-Embed & ICLR2025 & 4096 & 8B & 0.50 & 0.76 & 0.84 & 0.30 & 0.82 & 0.90 \\
    \cmidrule(lr){2-11}
    ~ & UniECS (Ours) & CIKM2025 & 256 & 0.2B & \textbf{0.62} & \textbf{0.93} & \textbf{0.98} & \textbf{0.41} & \textbf{0.87} & \textbf{0.95} \\ 
    \midrule
    \multirow{8}{*}{$(q_v,q_t) \to c_v$} & $\text{CLIP}_{SF}$ & ECCV2024 & 768 & 0.4B &  0.34 & 0.50 & 0.59 & 0.17 & 0.85 & 0.89 \\
    ~ & $\text{CLIP}_{FF}$ & ECCV2024 & 768 & 0.4B &  0.44 & 0.69 & 0.79 & 0.16 & 0.94 & 0.97 \\
    ~ & $\text{BLIP}_{SF}$ & ECCV2024 & 768 & 0.9B &  \textbf{0.65} & \underline{0.87} & \underline{0.93} & 0.15 & 0.94 & 0.97 \\
    ~ & $\text{BLIP}_{FF}$ & ECCV2024 & 768 & 0.9B &  0.41 & 0.69 & 0.80 & \underline{0.26} & 0.93 & 0.96 \\
    ~ & BGE-VL-large & - & 768 & 0.4B & 0.22 & 0.38 & 0.45 & 0.25 & 0.61 & 0.73 \\
    ~ & GME-Qwen2VL & CVPR2025 & 1536 & 2B & 0.50 & 0.75 & 0.82 & \textbf{0.27} & \underline{0.95} & \underline{0.98} \\
    ~ & MM-Embed & ICLR2025 & 4096 & 8B & 0.48 & 0.73 & 0.80 & 0.24 & 0.93 & 0.96 \\
    \cmidrule(lr){2-11}
    ~ & UniECS (Ours) & CIKM2025 & 256 & 0.2B & \underline{0.63} & \textbf{0.91} & \textbf{0.96} & 0.24 & \textbf{0.98} & \textbf{0.99} \\ 
    \midrule
    \multirow{8}{*}{$(q_v,q_t) \to (c_v,c_t)$} & $\text{CLIP}_{SF}$ & ECCV2024 & 768 & 0.4B &  0.51 & 0.74 & 0.82 & 0.13 & 0.81 & 0.86 \\
    ~ & $\text{CLIP}_{FF}$ & ECCV2024 & 768 & 0.4B &  0.56 & 0.82 & 0.89 & 0.26 & 0.94 & 0.97 \\
    ~ & $\text{BLIP}_{SF}$ & ECCV2024 & 768 & 0.9B &  0.66 & 0.88 & 0.94 & 0.28 & 0.95 & 0.98 \\
    ~ & $\text{BLIP}_{FF}$ & ECCV2024 & 768 & 0.9B &  0.52 & 0.77 & 0.86 & 0.30 & 0.96 & 0.98 \\
    ~ & BGE-VL-large & - & 768 & 0.4B & 0.25 & 0.42 & 0.52 & 0.28 & 0.71 & 0.79 \\
    ~ & GME-Qwen2VL & CVPR2025 & 1536 & 2B & \underline{0.71} & \underline{0.94} & \underline{0.98} & \underline{0.35} & \underline{0.97} & \underline{0.99} \\
    ~ & MM-Embed & ICLR2025 & 4096 & 8B & 0.58 & 0.85 & 0.92 & 0.23 & 0.93 & 0.96 \\
    \cmidrule(lr){2-11}
    ~ & UniECS (Ours) & CIKM2025 & 256 & 0.2B & \textbf{0.73} & \textbf{0.97} & \textbf{0.99} & \textbf{0.38} & \textbf{1.00} & \textbf{1.00} \\ 
    \bottomrule
\end{tabular*}
\end{adjustbox}
\label{tab:main-table-general-bench}
\end{table*}

In this section, we evaluate UniECS on multiple benchmarks to demonstrate its effectiveness for multimodal e-commerce retrieval. We first introduce datasets and experimental settings, followed by comparisons with baselines, embedding quality analysis, and ablation studies to analyze the contribution of each component. Additionally, we deploy our model in a production environment to measure its real-world impact.

\subsection{Experimental Setup}
\subsubsection{Datasets}
\par
For training our UniECS model, we construct a large-scale dataset of 10M product pairs by leveraging user interaction logs. 
We apply a filtering approach based on similarity by selecting product pairs where both visual similarity (computed by ViT) and textual similarity (computed by BERT) exceed a threshold of 0.75. 
This ensures genuine style and semantic relatedness. 
Each product appears only once to prevent information leakage. 

\begin{figure}
    \centering
    \includegraphics[width=0.8\linewidth]{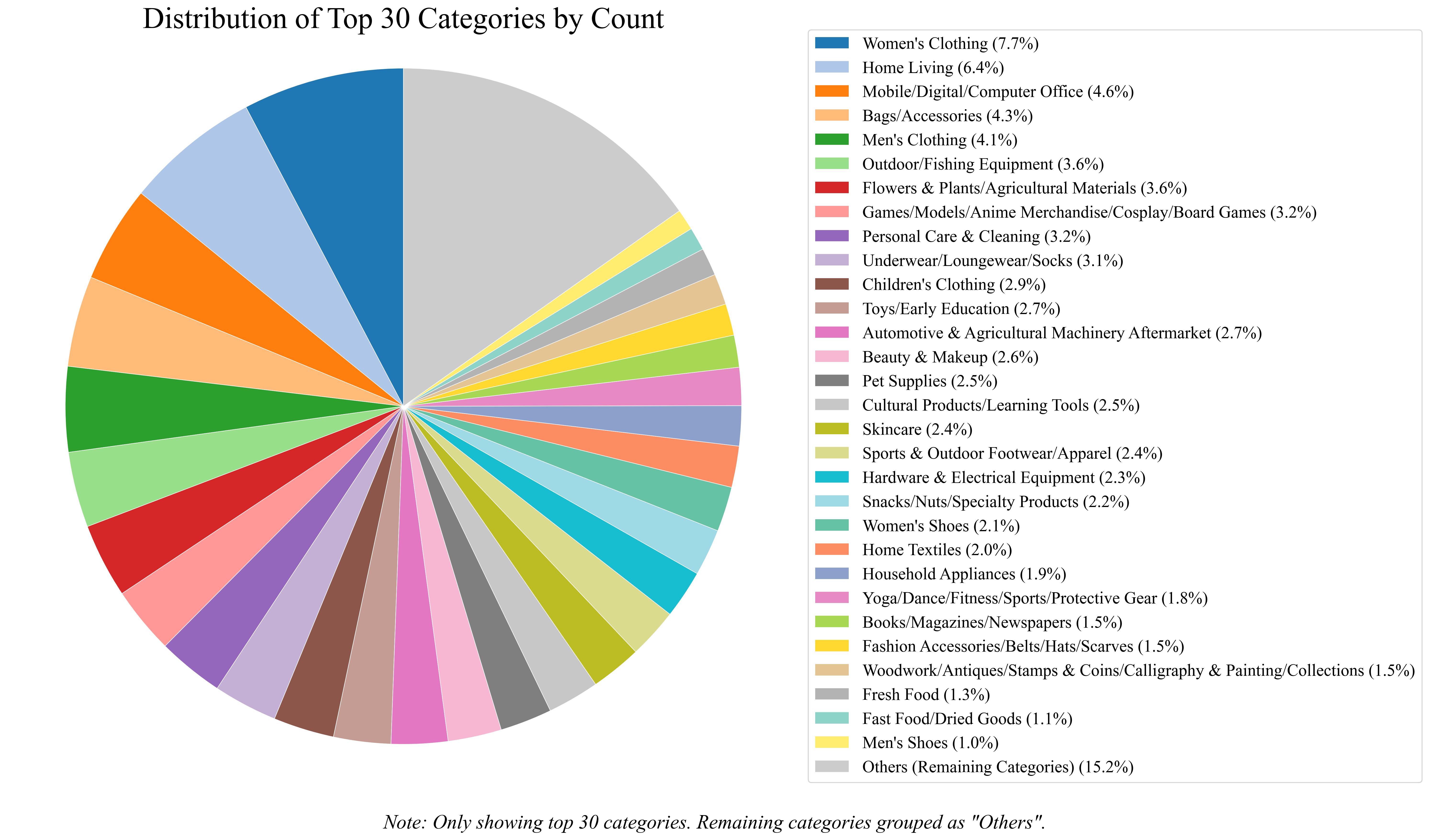}
    \caption{The category distribution of M-BEER dataset.}
    \label{fig:category_pie_chart}
\end{figure}

\par
For evaluation, we conduct experiments on four fashion datasets:

\textbf{M-BEER Bench}. 
We introduce the \textbf{M}ultimodal \textbf{BE}nchmark for \textbf{E}-Commerce \textbf{R}etrieval (M-BEER), which contains 50K product pairs for evaluating various retrieval scenarios. This benchmark is constructed using the same approach as our training data, with additional manual verification to ensure high-quality annotations. The category distribution of M-BEER is illustrated in Figure~\ref{fig:category_pie_chart}. Each sample in M-BEER consists of four key components: trigger text, trigger image, recall text, and recall image. This structure enables evaluation of all possible query-candidate combinations (e.g., text-to-image, image-to-text, multimodal-to-text) within the same benchmark, providing a standardized platform for comparing retrieval models across diverse e-commerce search scenarios.

\textbf{eSSPR Bench}~\cite{chen2023esspr}. 
An e-commerce search benchmark proposed by Alibaba that we evaluate in a zero-shot setting, and all models are tested directly without any fine-tuning. 
This allows us to assess the generalization capability of different models.

\textbf{Fashion200K}~\cite{Han2017fashion200k}. 
A fashion dataset containing 200K images with descriptions, which we use to evaluate cross-modal retrieval.

\textbf{FashionIQ}~\cite{wu2020fashioniqnewdataset}. 
A dataset containing fashion product images and descriptions of modifications, which we use to evaluate composed image retrieval performance.

\subsubsection{Evaluation Metrics}
We follow the standard retrieval evaluation metric, Recall@K, across nine different retrieval tasks covering all possible combinations of query and candidate modalities. Table~\ref{tab:retrieval_tasks} summarizes these tasks, where $q$ and $c$ denote query and candidate with subscripts $v$ (visual), $t$ (text), and $(v,t)$ (multimodal).

\begin{table}[!htbp]
\renewcommand{\arraystretch}{0.85}
\centering
\scriptsize
\caption{Retrieval tasks across modality combinations.}
\begin{tabular}{c|ccc}
\toprule
\textbf{Query} & \multicolumn{3}{c}{\textbf{Candidate Modality}} \\
\cmidrule{2-4}
\textbf{Modality} & $c_v$ & $c_t$ & $(c_v,c_t)$ \\
\midrule
$q_v$ & $q_v \rightarrow c_v$ & $q_v \rightarrow c_t$ & $q_v \rightarrow (c_v,c_t)$ \\
$q_t$ & $q_t \rightarrow c_v$ & $q_t \rightarrow c_t$ & $q_t \rightarrow (c_v,c_t)$ \\
$(q_v,q_t)$ & $(q_v,q_t) \rightarrow c_v$ & $(q_v,q_t) \rightarrow c_t$ & $(q_v,q_t) \rightarrow (c_v,c_t)$ \\
\bottomrule
\end{tabular}
\label{tab:retrieval_tasks}
\end{table}

\begin{figure*}[h]
    \centering
    \includegraphics[width=0.8\textwidth]{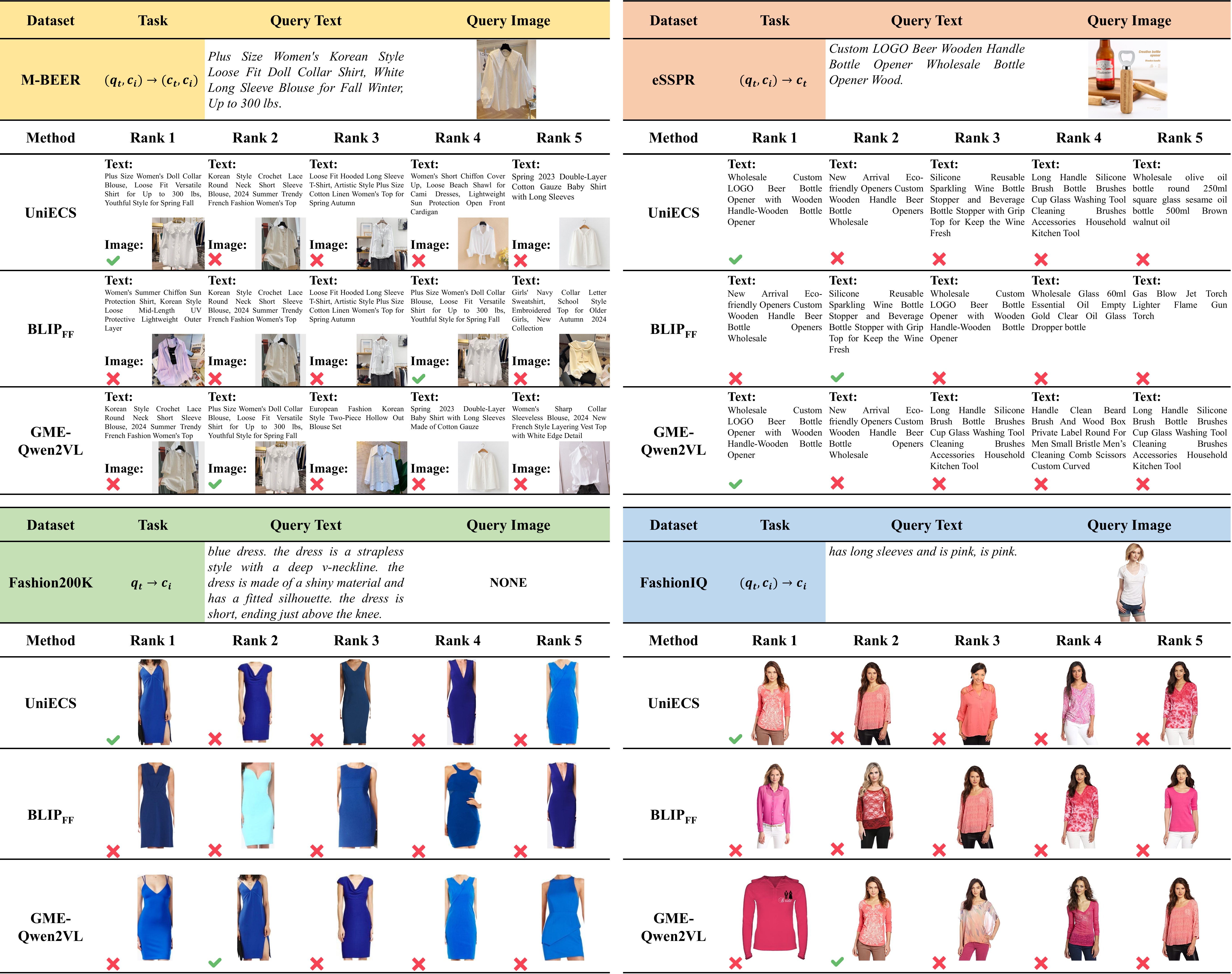}
    \caption{Visualization of top 5 retrieved candidates with three models. Check mark denotes the ground truth.}
    \label{fig:case}
\end{figure*}

\subsubsection{Baselines}
\label{sec:baselines}
We evaluate UniECS against a variety of models:
\begin{itemize}
\item \textbf{CLIP}~\cite{radford2021clip}: We evaluate two variants: ${\text{CLIP}_{SF}}$ (score fusion) and ${\text{CLIP}_{FF}}$ (feature fusion).
\item \textbf{BLIP}~\cite{li2022blip}: We include $\text{BLIP}_{SF}$ and $\text{BLIP}_{FF}$ variants.
\item \textbf{BGE-VL-large}~\cite{zhou2024bgevl}: A vision-language model based on CLIP and designed for multimodal retrieval.
\item \textbf{GME-Qwen2VL}~\cite{zhang2024gme}: A 2B-parameter multimodal embedding model based on Qwen2VL.
\item \textbf{MM-Embed}~\cite{lin2025mmembed}: An 8B-parameter multimodal embedding model proposed by Nvidia.
\end{itemize}

\subsubsection{Implementation Details}
Our visual encoder is a ViT-B/16 initialized from CLIP~\cite{radford2021clip}, and the text encoder is a BERT model with 12 layers.
For all representations (visual, textual, and multimodal features), we set the embedding dimension to 256. 
To train our model, we use the AdamW optimizer with a learning rate of $1\times10^{-4}$ and a batch size of 1024 distributed across 8 NVIDIA V100 GPUs. 
During training, we enhance the input images with data augmentation techniques including random cropping, horizontal flipping, and color jittering. 
The training process runs for 200K steps on a dataset of 10M product pairs.

For the hyperparameters, we set $\alpha_1=0.2$, $\alpha_2=0.1$, $\alpha_3=0.05$, and $\alpha_4=0.2$. 
The temperature parameters $\tau$, $\tau_v$, and $\tau_t$ are set to 0.07, 0.07, and 0.03, respectively. 
The loss weights are initially set to 1.0 and dynamically adjusted during training using our adaptive weighting scheme with $\beta=0.5$. We validate the choice of $\beta=0.5$ through systematic ablation studies presented in Section 4.3.3.

\begin{table}[htbp]
\centering
\scriptsize
\renewcommand{\arraystretch}{0.85}
\caption{Benchmarking performance on Fashion200K and FashionIQ. \textbf{Bold}: top-1 performance. \underline{Underline}: top-2.}
\begin{adjustbox}{max width=0.875\linewidth}
\begin{tabular*}{0.85\linewidth}{@{\extracolsep{\fill}}c|ccc}
\toprule
\textbf{Method} & \textbf{R@1} & \textbf{R@5} & \textbf{R@10} \\
\midrule
\rowcolor{lightgray}
\multicolumn{4}{c}{Fashion200K / $q_t \to c_v$} \\
\midrule
CLIP$_{SF}$ & 0.05 & 0.14 & 0.18 \\
CLIP$_{FF}$ & 0.04 & 0.13 & 0.16 \\
BLIP$_{SF}$ & 0.07 & 0.19 & 0.24 \\
BLIP$_{FF}$ & 0.07 & \underline{0.21} & \underline{0.25} \\
BGE-VL-large & 0.06 & 0.19 & 0.23 \\
GME-Qwen2VL & \underline{0.08} & 0.20 & \underline{0.25}  \\
MM-Embed & 0.05 & 0.13 & 0.17 \\
UniECS (Ours) & \textbf{0.09} & \textbf{0.24} & \textbf{0.31} \\
\midrule
\rowcolor{lightgray}
\multicolumn{4}{c}{Fashion200K / $q_v \to c_t$} \\
\midrule
CLIP$_{SF}$ & 0.07 & 0.14 & 0.18 \\
CLIP$_{FF}$ & 0.05 & 0.11 & 0.14 \\
BLIP$_{SF}$ & 0.10 & 0.19 & 0.24 \\
BLIP$_{FF}$ & \underline{0.11} & \underline{0.20} & \underline{0.26} \\
BGE-VL-large & 0.08 & 0.16 & 0.20 \\
GME-Qwen2VL & 0.10 & \textbf{0.21} & \textbf{0.27} \\
MM-Embed & 0.06 & 0.13 & 0.18 \\
UniECS (Ours) & \textbf{0.12} & 0.19 & \underline{0.26} \\
\midrule
\rowcolor{lightgray}
\multicolumn{4}{c}{FashionIQ / $(q_v,q_t) \to c_v$} \\
\midrule
CLIP$_{SF}$ & 0.10 & 0.19 & 0.24 \\
CLIP$_{FF}$ & 0.07 & 0.17 & 0.21 \\
BLIP$_{SF}$ & 0.11 & 0.21 & 0.26 \\
BLIP$_{FF}$ & 0.12 & 0.23 & 0.29 \\
BGE-VL-large & \underline{0.14} & \underline{0.28} & \underline{0.35} \\
GME-Qwen2VL & 0.09 & 0.18 & 0.22 \\
MM-Embed & 0.14 & 0.27 & 0.33 \\
UniECS (Ours) & \textbf{0.19} & \textbf{0.36} & \textbf{0.49} \\
\bottomrule
\end{tabular*}
\end{adjustbox}
\label{tab:fashion-datasets}
\end{table}

\subsection{Experimental Results}
\par
We present comprehensive experimental results of UniECS on four fashion datasets, analyze its performance across different retrieval tasks, and examine the quality of learned embeddings.

\subsubsection{Results on Comprehensive Fashion Benchmarks}

\par
Table~\ref{tab:main-table-general-bench} presents the retrieval results on M-BEER Bench and eSSPR Bench.
On M-BEER Bench, UniECS outperforms baselines in most retrieval tasks, achieving obvious improvements especially in cross-modal tasks.

\par
For single-modality retrieval tasks, UniECS demonstrates consistently strong performance. 
In image-to-image retrieval ($q_v \rightarrow c_v$), our model achieves 0.95 R@10 on M-BEER Bench, surpassing BLIP$_{SF}$ (0.94 R@10) while also delivering competitive 0.63 R@1. 
For text-to-text retrieval ($q_t \rightarrow c_t$), UniECS reaches 0.98 R@10, outperforming GME-Qwen2VL (0.95 R@10). 
Most notably, in multimodal to multimodal retrieval ($(q_v,q_t) \rightarrow (c_v,c_t)$), UniECS achieves near-perfect performance with 0.99 R@10, surpassing GME-Qwen2VL (0.98 R@10) and demonstrating the effectiveness.

\par
The most obvious improvements are observed in cross-modal retrieval tasks. 
For text-to-image retrieval ($q_t \rightarrow c_v$), UniECS achieves 0.85 R@10, substantially outperforming GME-Qwen2VL (0.57 R@10) by 28 percentage points, while also showing impressive gains at R@1 (0.36 vs. 0.26). 
Similarly, for image-to-text retrieval ($q_v \rightarrow c_t$), UniECS achieves 0.87 R@10, a remarkable 27 percentage point improvement over GME-Qwen2VL (0.60 R@10). 
Other cross-modal combinations also show notable gains in R@10, such as image-to-multimodal ($q_v \rightarrow (c_v,c_t)$) with 0.97 R@10 versus 0.85 for GME-Qwen2VL, and multimodal-to-image ($(q_v,q_t) \rightarrow c_v$) with 0.96 R@10 versus 0.82 for GME. 
These consistent improvements confirm that UniECS effectively aligns different modal spaces.

\par
For zero-shot evaluation on eSSPR Bench, UniECS demonstrates strong generalization capability by achieving competitive or superior performance compared to far larger models.
In the text-to-text retrieval task, UniECS achieves an impressive 0.98 R@10, outperforming both GME-Qwen2VL (0.92 R@10) and MM-Embed (0.89 R@10) despite having only a fraction of their parameter count. Similarly, for multimodal queries to text candidates ($(q_v,q_t) \rightarrow c_t$), UniECS achieves 0.95 R@10, outperforming GME-Qwen2VL with 0.94 R@10, further demonstrating its strong cross-modal capabilities in zero-shot settings.

\par
The consistent superior performance across all retrieval tasks demonstrates the effectiveness of our gated multimodal encoder and comprehensive loss design in capturing both modality-specific and cross-modal information. 
Most importantly, UniECS achieves these results with only 0.2B parameters and 256-dimensional embeddings, making it more efficient than competing methods such as GME-Qwen2VL (2B, 1536-dim) and MM-Embed (8B, 4096-dim), which is crucial for real-world applications requiring low-latency.

\subsubsection{Results on Task-specific Fashion Benchmarks}
\par
Table~\ref{tab:fashion-datasets} shows the retrieval results on Fashion200K and FashionIQ. 
On Fashion200K, which focuses on text-to-image and image-to-text retrieval, UniECS displays consistent advantages over baselines. 
For text-to-image retrieval ($q_t \rightarrow c_v$), UniECS achieves 0.31 R@10, outperforming BLIP$_{FF}$ and GME-Qwen2VL (both 0.25 R@10) by 6 percentage points. 
While the absolute performance are lower than on M-BEER Bench due to the challenging nature of the dataset, the relative improvements are substantial.

\par
On FashionIQ ($(q_v,q_t) \rightarrow c_v$), UniECS also outperform all baselines, achieving 0.49 R@10, which is 14 percentage points higher than BGE-VL-large and MM-Embed (0.35 and 0.33 R@10, respectively). 
This improvement suggests our framework effectively combines visual and textual information in fashion retrieval. 
Such composition capabilities, where users provide both an image and modification text, are highly practical for e-commerce applications.

\par
Figure~\ref{fig:case} provides qualitative examples of retrieval results across four datasets, comparing our UniECS against BLIP$_{FF}$ and GME-Qwen2VL. The visualization shows UniECS consistently retrieves more relevant products.

\subsubsection{Embedding Quality Analysis}
\par
\begin{figure}[!t]
    \centering
    \includegraphics[width=0.8\linewidth]{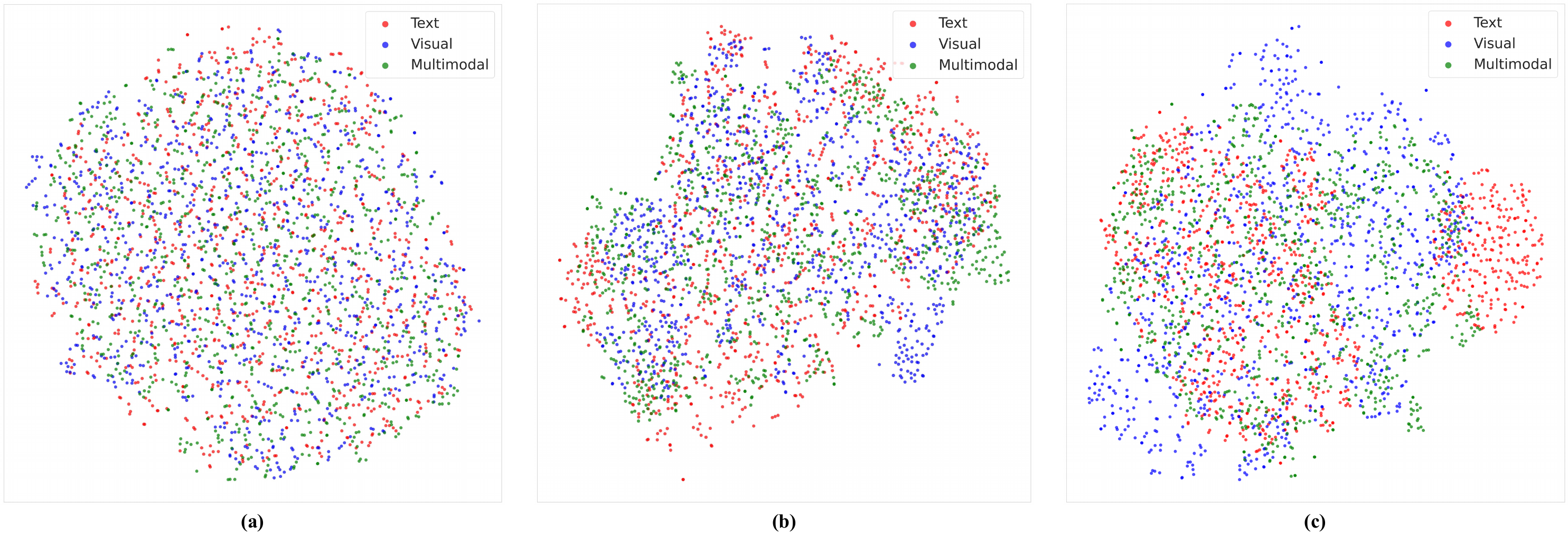}
    \caption{Visualization of the embeddings in a 2D plot by T-SNE. (a): Our UniECS, (b): GME-Qwen2VL, (c): BLIP$_{FF}$.}
   
    \label{fig:emb}
\end{figure}

To understand the effectiveness of our unified framework, we analyze whether our model produces modality-universal embeddings where semantically similar content from different modalities clusters together.
We sample 1000 product instances from M-BEER benchmark and visualize their embeddings using t-SNE, as shown in Figure~\ref{fig:emb}.
BLIP$_{FF}$ shows modality separation with separate clusters, GME-Qwen2VL exhibits moderate cross-modal mixing, while UniECS demonstrates the most uniform distribution organized by semantic content rather than modality type.
This pattern confirms that our gated multimodal encoder successfully creates a unified embedding space where semantically similar products cluster together regardless of input modality.

\subsection{Ablation Studies}
\par
To analyze component contributions and hyperparameter sensitivity, we conduct ablation studies on M-BEER Bench. Results are presented in Table \ref{tab:ablation} and Figure~\ref{fig:adaptive_loss}.

\subsubsection{Effect of Loss Groups}
We examine the impact of three loss groups on UniECS:
\begin{itemize}
    \item \textbf{Cross-Modal Alignment Loss (CMAL)}: Removing CMAL leads to a noticeable drop in performance across all retrieval tasks, with clear impact on cross-modal tasks. 
    For example, R@1 drops by 11 percentage points on $q_t \rightarrow c_v$ and 12 percentage points on $q_v \rightarrow c_t$. 
    This confirms the importance of aligning different modalities.
    
    \item \textbf{Cohesive Local Alignment Loss (CLAL)}: When CLAL is removed, performance decreases moderately across all tasks. 
    This indicates that these regularization terms help maintain consistent feature structures across different modalities and improve the quality of the learned representations.
    
    \item \textbf{Intra-Modal Contrastive Loss (IMCL)}: Removing IMCL results in notable performance drops in single-modality retrieval tasks.
    R@1 on $q_t \rightarrow c_t$ drops from 0.62 to 0.48, highlighting the importance of enhancing modality-specific representations.
\end{itemize}

\subsubsection{Impact of Gating Mechanism}
\par
The adaptive gating mechanism proves to be critical for handling missing modalities and balancing the contribution of each modality. 
When the gating mechanism is removed, performance drops drastically on image-related tasks.
For example, R@1 on $q_v \rightarrow c_v$ drops drastically from 0.63 to 0.12, a 51 percentage point decrease.

\begin{table}[!htbp]
\scriptsize
\centering
\caption{Ablation study on different loss functions and gating mechanism on M-BEER Bench.}
\renewcommand{\arraystretch}{0.85}
\setlength{\tabcolsep}{1.5pt}
\begin{tabular}{c|ccc|ccc|ccc}
\toprule
\multirow{2}{*}{\textbf{Method}} & \multicolumn{3}{c|}{$q_t \rightarrow c_v$} & \multicolumn{3}{c|}{$q_t \rightarrow c_t$} & \multicolumn{3}{c}{$q_t \rightarrow (c_v,c_t)$} \\
\cmidrule{2-10}
 & \textbf{R@1} & \textbf{R@5} & \textbf{R@10} & \textbf{R@1} & \textbf{R@5} & \textbf{R@10} & \textbf{R@1} & \textbf{R@5} & \textbf{R@10} \\
\midrule
UniECS (Full) & \textbf{0.36} & \textbf{0.74} & \textbf{0.85} & \underline{0.62} & \textbf{0.93} & \textbf{0.98} & \textbf{0.63} & \textbf{0.94} & \textbf{0.98} \\
w/o CMAL & 0.25 & 0.58 & 0.70 & 0.57 & 0.86 & 0.93 & 0.47 & 0.80 & 0.89 \\
w/o CLAL & \underline{0.33} & \underline{0.69} & \underline{0.81} & 0.59 & \underline{0.91} & \underline{0.96} & 0.60 & 0.92 & \underline{0.97} \\
w/o IMCL & 0.32 & 0.68 & 0.80 & 0.48 & 0.80 & 0.88 & 0.58 & 0.90 & 0.95 \\
w/o Gating & 0.18 & 0.42 & 0.53 & \textbf{0.63} & \textbf{0.93} & \textbf{0.98} & \underline{0.62} & \underline{0.93} & \textbf{0.98} \\
\midrule
\midrule
\multirow{2}{*}{\textbf{Method}} & \multicolumn{3}{c|}{$q_v \rightarrow c_t$} & \multicolumn{3}{c|}{$q_v \rightarrow c_v$} & \multicolumn{3}{c}{$q_v \rightarrow (c_v,c_t)$} \\
\cmidrule{2-10}
 & \textbf{R@1} & \textbf{R@5} & \textbf{R@10} & \textbf{R@1} & \textbf{R@5} & \textbf{R@10} & \textbf{R@1} & \textbf{R@5} & \textbf{R@10} \\
\midrule
UniECS (Full) & \textbf{0.38} & \textbf{0.77} & \textbf{0.87} & \textbf{0.63} & \textbf{0.89} & \textbf{0.95} & \textbf{0.64} & \textbf{0.92} & \textbf{0.97} \\
w/o CMAL & 0.26 & 0.61 & 0.73 & 0.56 & 0.84 & 0.92 & 0.52 & 0.81 & 0.87 \\
w/o CLAL & \underline{0.35} & \underline{0.73} & \underline{0.84} & \underline{0.61} & \underline{0.87} & \underline{0.94} & \underline{0.61} & \underline{0.89} & \underline{0.95} \\
w/o IMCL & 0.32 & 0.71 & 0.80 & 0.52 & 0.81 & 0.89 & 0.56 & 0.86 & 0.92 \\
w/o Gating & 0.15 & 0.36 & 0.47 & 0.12 & 0.30 & 0.42 & 0.14 & 0.33 & 0.45 \\
\midrule
\midrule
\multirow{2}{*}{\textbf{Method}} & \multicolumn{3}{c|}{$(q_v,q_t) \rightarrow c_t$} & \multicolumn{3}{c|}{$(q_v,q_t) \rightarrow c_v$} & \multicolumn{3}{c}{$(q_v,q_t) \rightarrow (c_v,c_t)$} \\
\cmidrule{2-10}
 & \textbf{R@1} & \textbf{R@5} & \textbf{R@10} & \textbf{R@1} & \textbf{R@5} & \textbf{R@10} & \textbf{R@1} & \textbf{R@5} & \textbf{R@10} \\
\midrule
UniECS (Full) & \textbf{0.62} & \underline{0.93} & \textbf{0.98} & \textbf{0.63} & \textbf{0.91} & \textbf{0.96} & \textbf{0.73} & \textbf{0.97} & \textbf{0.99} \\
w/o CMAL & 0.48 & 0.81 & 0.87 & 0.51 & 0.83 & 0.88 & 0.58 & 0.86 & 0.92 \\
w/o CLAL & 0.58 & 0.90 & \underline{0.96} & \underline{0.61} & \underline{0.89} & \underline{0.95} & 0.70 & 0.94 & \underline{0.98} \\
w/o IMCL & 0.56 & 0.87 & 0.93 & 0.57 & 0.87 & 0.93 & 0.67 & 0.92 & 0.97 \\
w/o Gating & \underline{0.61} & \textbf{0.94} & \textbf{0.98} & 0.14 & 0.35 & 0.46 & \underline{0.71} & \underline{0.95} & 0.97 \\
\bottomrule
\end{tabular}
\label{tab:ablation}
\end{table}

\par
Interestingly, the impact on text-only retrieval ($q_t \rightarrow c_t$) is minimal when gating is removed. 
This phenomenon reveals that without the gating mechanism, the text modality dominates the fusion process, leading to poor visual representation performance. 
The adaptive gating mechanism effectively balances modality contributions, preventing text features from overwhelming visual features.

\subsubsection{Impact of Adaptive Loss Weighting}
\par
\begin{figure}[h]
    \centering
    \includegraphics[width=0.8\linewidth]{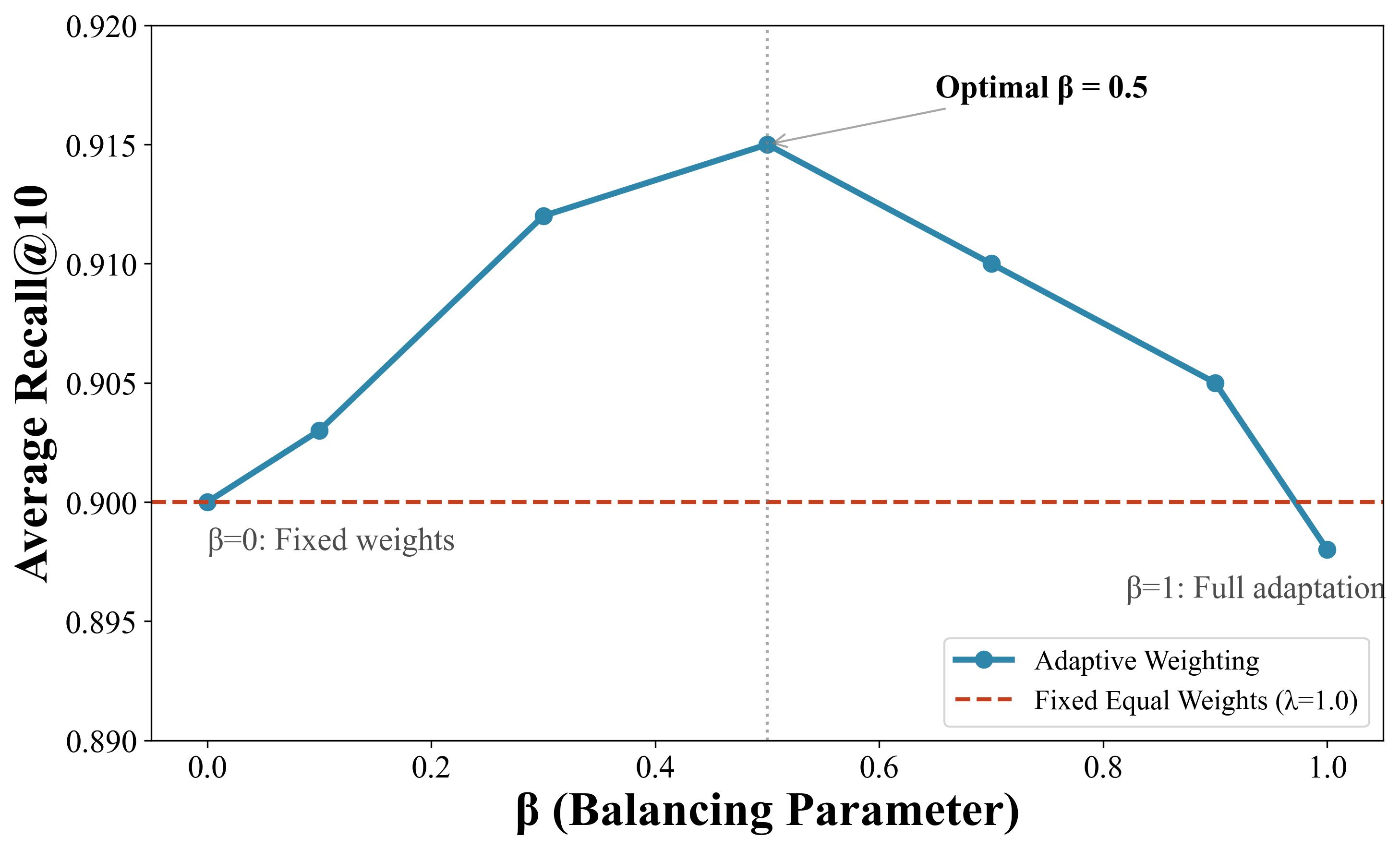}
    \caption{Ablation study of the parameter $\beta$.}
    \label{fig:adaptive_loss}
   
\end{figure}

We examine the effect of the balancing parameter $\beta$ in our adaptive loss weighting scheme on M-BEER Bench. 
Figure~\ref{fig:adaptive_loss} shows the performance under different $\beta$ values, where $\beta = 0$ corresponds to fixed equal weights (red baseline). 
The results demonstrate that $\beta = 0.5$ achieves optimal performance, validating our hyperparameter choice.

\subsection{Online Experiments}
We deploy UniECS in the e-commerce search platform of Kuaishou Inc. across two scenarios, image search  and same-style item retrieval. The first only take the user's image as input, and return the item list with similar image. While the second scenario takes the item (image + text descriptions) as input, and outputs the same style items. Performance is evaluated through A/B testing comparing key business metrics before and after deployment. As shown in Table~\ref{tab:online}, the deployment achieves notable improvements across all measured metrics, with Same-style Coverage Rate (SCR) increased by 27.31\%, Click-Through Rate (CTR) by 2.74\%, Average Conversion Rate (CVR) by a 4.67\%, Order volume by 6.12\%, and total revenue increased 8.33\% on average. These all indicate our method can substantially improve the user search experience, and ultimately boost the industry revenue.

\begin{table}[h]
\centering
\scriptsize
\renewcommand{\arraystretch}{0.85}
\caption{Online results for A/B testing.}
\begin{adjustbox}{max width=\linewidth}
\begin{tabular*}{0.9\linewidth}{@{\extracolsep{\fill}}cccccc@{\extracolsep{\fill}}}
\toprule
\textbf{Scenario} & \textbf{SCR} & \textbf{CTR} & \textbf{CVR} & \textbf{Order} & \textbf{Revenue} \\
\midrule
Image Retrieval & 28.50\% & 2.059\% & 5.01\% & 7.46\% & 5.47\% \\
Same-style Retrieval & 26.13\% & 3.437\% & 4.33\% & 4.78\% &  11.20\%\\
\bottomrule
\end{tabular*}
\end{adjustbox}
\label{tab:online}
\end{table}

To facilitate future research, corresponding codes, models and datasets will be made publicly available.

\section{Conclusion}
\par
We introduce UniECS, a unified multimodal e-commerce search framework that handles visual, textual, and combined inputs through three key innovations: a flexible architecture supporting various modality combinations, an adaptive gating mechanism for missing modalities, and a comprehensive loss design enhancing cross-modal alignment.
Experiments on multiple benchmarks, including our M-BEER dataset with 50K product pairs, show UniECS consistently outperforms existing methods while maintaining parameter efficiency (0.2B vs. larger models like GME-Qwen2VL and MM-Embed).

\section{GenAI Usage Disclosure}
\par
We hereby declare that throughout the innovative conceptualization, code development, data construction, manuscript preparation, and creation of all figures and tables in this research, \textbf{no artificial intelligence generation tools were utilized}.

\bibliographystyle{ACM-Reference-Format}
\bibliography{reference}


\begin{thebibliography}{33}


\ifx \showCODEN    \undefined \def \showCODEN     #1{\unskip}     \fi
\ifx \showISBNx    \undefined \def \showISBNx     #1{\unskip}     \fi
\ifx \showISBNxiii \undefined \def \showISBNxiii  #1{\unskip}     \fi
\ifx \showISSN     \undefined \def \showISSN      #1{\unskip}     \fi
\ifx \showLCCN     \undefined \def \showLCCN      #1{\unskip}     \fi
\ifx \shownote     \undefined \def \shownote      #1{#1}          \fi
\ifx \showarticletitle \undefined \def \showarticletitle #1{#1}   \fi
\ifx \showURL      \undefined \def \showURL       {\relax}        \fi
\providecommand\bibfield[2]{#2}
\providecommand\bibinfo[2]{#2}
\providecommand\natexlab[1]{#1}
\providecommand\showeprint[2][]{arXiv:#2}

\bibitem[Agrawal et~al\mbox{.}(2024)]%
        {agrawal2024pixtral12b}
\bibfield{author}{\bibinfo{person}{Pravesh Agrawal}, \bibinfo{person}{Szymon Antoniak}, \bibinfo{person}{Emma~Bou Hanna}, \bibinfo{person}{Baptiste Bout}, \bibinfo{person}{Devendra Chaplot}, \bibinfo{person}{Jessica Chudnovsky}, \bibinfo{person}{Diogo Costa}, \bibinfo{person}{Baudouin~De Monicault}, \bibinfo{person}{Saurabh Garg}, \bibinfo{person}{Theophile Gervet}, \bibinfo{person}{Soham Ghosh}, \bibinfo{person}{Amélie Héliou}, \bibinfo{person}{Paul Jacob}, \bibinfo{person}{Albert~Q. Jiang}, \bibinfo{person}{Kartik Khandelwal}, \bibinfo{person}{Timothée Lacroix}, \bibinfo{person}{Guillaume Lample}, \bibinfo{person}{Diego~Las Casas}, \bibinfo{person}{Thibaut Lavril}, \bibinfo{person}{Teven~Le Scao}, \bibinfo{person}{Andy Lo}, \bibinfo{person}{William Marshall}, \bibinfo{person}{Louis Martin}, \bibinfo{person}{Arthur Mensch}, \bibinfo{person}{Pavankumar Muddireddy}, \bibinfo{person}{Valera Nemychnikova}, \bibinfo{person}{Marie Pellat}, \bibinfo{person}{Patrick~Von Platen}, \bibinfo{person}{Nikhil
  Raghuraman}, \bibinfo{person}{Baptiste Rozière}, \bibinfo{person}{Alexandre Sablayrolles}, \bibinfo{person}{Lucile Saulnier}, \bibinfo{person}{Romain Sauvestre}, \bibinfo{person}{Wendy Shang}, \bibinfo{person}{Roman Soletskyi}, \bibinfo{person}{Lawrence Stewart}, \bibinfo{person}{Pierre Stock}, \bibinfo{person}{Joachim Studnia}, \bibinfo{person}{Sandeep Subramanian}, \bibinfo{person}{Sagar Vaze}, \bibinfo{person}{Thomas Wang}, {and} \bibinfo{person}{Sophia Yang}.} \bibinfo{year}{2024}\natexlab{}.
\newblock \bibinfo{title}{Pixtral 12B}.
\newblock
\showeprint[arxiv]{2410.07073}~[cs.CV]


\bibitem[Alayrac et~al\mbox{.}(2022)]%
        {alayrac2022flamingo}
\bibfield{author}{\bibinfo{person}{Jean-Baptiste Alayrac}, \bibinfo{person}{Jeff Donahue}, \bibinfo{person}{Pauline Luc}, \bibinfo{person}{Antoine Miech}, \bibinfo{person}{Iain Barr}, \bibinfo{person}{Yana Hasson}, \bibinfo{person}{Karel Lenc}, \bibinfo{person}{Arthur Mensch}, \bibinfo{person}{Katie Millican}, \bibinfo{person}{Malcolm Reynolds}, \bibinfo{person}{Roman Ring}, \bibinfo{person}{Eliza Rutherford}, \bibinfo{person}{Serkan Cabi}, \bibinfo{person}{Tengda Han}, \bibinfo{person}{Zhitao Gong}, \bibinfo{person}{Sina Samangooei}, \bibinfo{person}{Marianne Monteiro}, \bibinfo{person}{Jacob Menick}, \bibinfo{person}{Sebastian Borgeaud}, \bibinfo{person}{Andrew Brock}, \bibinfo{person}{Aida Nematzadeh}, \bibinfo{person}{Sahand Sharifzadeh}, \bibinfo{person}{Mikolaj Binkowski}, \bibinfo{person}{Ricardo Barreira}, \bibinfo{person}{Oriol Vinyals}, \bibinfo{person}{Andrew Zisserman}, {and} \bibinfo{person}{Karen Simonyan}.} \bibinfo{year}{2022}\natexlab{}.
\newblock \bibinfo{title}{Flamingo: a Visual Language Model for Few-Shot Learning}.
\newblock
\showeprint[arxiv]{2204.14198}~[cs.CV]


\bibitem[Chen et~al\mbox{.}(2023)]%
        {chen2023esspr}
\bibfield{author}{\bibinfo{person}{Ben Chen}, \bibinfo{person}{Linbo Jin}, \bibinfo{person}{Xinxin Wang}, \bibinfo{person}{Dehong Gao}, \bibinfo{person}{Wen Jiang}, {and} \bibinfo{person}{Wei Ning}.} \bibinfo{year}{2023}\natexlab{}.
\newblock \showarticletitle{Unified Vision-Language Representation Modeling for E-Commerce Same-style Products Retrieval}. In \bibinfo{booktitle}{\emph{Companion Proceedings of the ACM Web Conference 2023}} (Austin, TX, USA). \bibinfo{pages}{381–385}.
\newblock


\bibitem[Dai et~al\mbox{.}(2024)]%
        {dai2024nvlmopenfrontierclassmultimodal}
\bibfield{author}{\bibinfo{person}{Wenliang Dai}, \bibinfo{person}{Nayeon Lee}, \bibinfo{person}{Boxin Wang}, \bibinfo{person}{Zhuolin Yang}, \bibinfo{person}{Zihan Liu}, \bibinfo{person}{Jon Barker}, \bibinfo{person}{Tuomas Rintamaki}, \bibinfo{person}{Mohammad Shoeybi}, \bibinfo{person}{Bryan Catanzaro}, {and} \bibinfo{person}{Wei Ping}.} \bibinfo{year}{2024}\natexlab{}.
\newblock \bibinfo{title}{NVLM: Open Frontier-Class Multimodal LLMs}.
\newblock
\showeprint[arxiv]{2409.11402}~[cs.CL]


\bibitem[Deng et~al\mbox{.}(2023)]%
        {Deng2023MIND2WEB}
\bibfield{author}{\bibinfo{person}{Xiang Deng}, \bibinfo{person}{Yu Gu}, \bibinfo{person}{Boyuan Zheng}, \bibinfo{person}{Shijie Chen}, \bibinfo{person}{Samuel Stevens}, \bibinfo{person}{Boshi Wang}, \bibinfo{person}{Huan Sun}, {and} \bibinfo{person}{Yu Su}.} \bibinfo{year}{2023}\natexlab{}.
\newblock \showarticletitle{MIND2WEB: towards a generalist agent for the web}. In \bibinfo{booktitle}{\emph{Proceedings of the 37th International Conference on Neural Information Processing Systems}} (New Orleans, LA, USA) \emph{(\bibinfo{series}{NIPS '23})}.
\newblock


\bibitem[Hadsell et~al\mbox{.}(2006)]%
        {Raia2006contrastiveloss}
\bibfield{author}{\bibinfo{person}{Raia Hadsell}, \bibinfo{person}{Sumit Chopra}, {and} \bibinfo{person}{Yann LeCun}.} \bibinfo{year}{2006}\natexlab{}.
\newblock \showarticletitle{Dimensionality Reduction by Learning an Invariant Mapping}. In \bibinfo{booktitle}{\emph{Proceedings of the 2006 IEEE Computer Society Conference on Computer Vision and Pattern Recognition - Volume 2}}. \bibinfo{address}{USA}, \bibinfo{pages}{1735–1742}.
\newblock


\bibitem[Han et~al\mbox{.}(2017)]%
        {Han2017fashion200k}
\bibfield{author}{\bibinfo{person}{Xintong Han}, \bibinfo{person}{Zuxuan Wu}, \bibinfo{person}{Phoenix~X. Huang}, \bibinfo{person}{Xiao Zhang}, \bibinfo{person}{Menglong Zhu}, \bibinfo{person}{Yuan Li}, \bibinfo{person}{Yang Zhao}, {and} \bibinfo{person}{Larry~S. Davis}.} \bibinfo{year}{2017}\natexlab{}.
\newblock \showarticletitle{Automatic Spatially-Aware Fashion Concept Discovery}. In \bibinfo{booktitle}{\emph{2017 IEEE International Conference on Computer Vision (ICCV)}}. \bibinfo{pages}{1472--1480}.
\newblock


\bibitem[Hong et~al\mbox{.}(2024)]%
        {hong2024metagpt}
\bibfield{author}{\bibinfo{person}{Sirui Hong}, \bibinfo{person}{Mingchen Zhuge}, \bibinfo{person}{Jonathan Chen}, \bibinfo{person}{Xiawu Zheng}, \bibinfo{person}{Yuheng Cheng}, \bibinfo{person}{Jinlin Wang}, \bibinfo{person}{Ceyao Zhang}, \bibinfo{person}{Zili Wang}, \bibinfo{person}{Steven Ka~Shing Yau}, \bibinfo{person}{Zijuan Lin}, \bibinfo{person}{Liyang Zhou}, \bibinfo{person}{Chenyu Ran}, \bibinfo{person}{Lingfeng Xiao}, \bibinfo{person}{Chenglin Wu}, {and} \bibinfo{person}{J{\"u}rgen Schmidhuber}.} \bibinfo{year}{2024}\natexlab{}.
\newblock \showarticletitle{Meta{GPT}: Meta Programming for A Multi-Agent Collaborative Framework}. In \bibinfo{booktitle}{\emph{The Twelfth International Conference on Learning Representations}}.
\newblock


\bibitem[Jia et~al\mbox{.}(2021)]%
        {jia2021align}
\bibfield{author}{\bibinfo{person}{Chao Jia}, \bibinfo{person}{Yinfei Yang}, \bibinfo{person}{Ye Xia}, \bibinfo{person}{Yi-Ting Chen}, \bibinfo{person}{Zarana Parekh}, \bibinfo{person}{Hieu Pham}, \bibinfo{person}{Quoc~V. Le}, \bibinfo{person}{Yunhsuan Sung}, \bibinfo{person}{Zhen Li}, {and} \bibinfo{person}{Tom Duerig}.} \bibinfo{year}{2021}\natexlab{}.
\newblock \bibinfo{title}{Scaling Up Visual and Vision-Language Representation Learning With Noisy Text Supervision}.
\newblock
\showeprint[arxiv]{2102.05918}~[cs.CV]


\bibitem[Jiang et~al\mbox{.}(2024)]%
        {jiang2024e5v}
\bibfield{author}{\bibinfo{person}{Ting Jiang}, \bibinfo{person}{Minghui Song}, \bibinfo{person}{Zihan Zhang}, \bibinfo{person}{Haizhen Huang}, \bibinfo{person}{Weiwei Deng}, \bibinfo{person}{Feng Sun}, \bibinfo{person}{Qi Zhang}, \bibinfo{person}{Deqing Wang}, {and} \bibinfo{person}{Fuzhen Zhuang}.} \bibinfo{year}{2024}\natexlab{}.
\newblock \bibinfo{title}{E5-V: Universal Embeddings with Multimodal Large Language Models}.
\newblock
\showeprint[arxiv]{2407.12580}~[cs.CL]


\bibitem[Kim et~al\mbox{.}(2020)]%
        {Kim2020ProxyAnchorLoss}
\bibfield{author}{\bibinfo{person}{Sungyeon Kim}, \bibinfo{person}{Dongwon Kim}, \bibinfo{person}{Minsu Cho}, {and} \bibinfo{person}{Suha Kwak}.} \bibinfo{year}{2020}\natexlab{}.
\newblock \showarticletitle{Proxy Anchor Loss for Deep Metric Learning}. In \bibinfo{booktitle}{\emph{2020 IEEE/CVF Conference on Computer Vision and Pattern Recognition (CVPR)}}. \bibinfo{pages}{3235--3244}.
\newblock


\bibitem[Kim et~al\mbox{.}(2024)]%
        {kim2024mdagents}
\bibfield{author}{\bibinfo{person}{Yubin Kim}, \bibinfo{person}{Chanwoo Park}, \bibinfo{person}{Hyewon Jeong}, \bibinfo{person}{Yik~Siu Chan}, \bibinfo{person}{Xuhai Xu}, \bibinfo{person}{Daniel McDuff}, \bibinfo{person}{Hyeonhoon Lee}, \bibinfo{person}{Marzyeh Ghassemi}, \bibinfo{person}{Cynthia Breazeal}, {and} \bibinfo{person}{Hae~Won Park}.} \bibinfo{year}{2024}\natexlab{}.
\newblock \showarticletitle{{MDA}gents: An Adaptive Collaboration of {LLM}s for Medical Decision-Making}. In \bibinfo{booktitle}{\emph{The Thirty-eighth Annual Conference on Neural Information Processing Systems}}.
\newblock


\bibitem[Li et~al\mbox{.}(2023)]%
        {li2023blip2}
\bibfield{author}{\bibinfo{person}{Junnan Li}, \bibinfo{person}{Dongxu Li}, \bibinfo{person}{Silvio Savarese}, {and} \bibinfo{person}{Steven Hoi}.} \bibinfo{year}{2023}\natexlab{}.
\newblock \showarticletitle{{BLIP}-2: Bootstrapping Language-Image Pre-training with Frozen Image Encoders and Large Language Models}. In \bibinfo{booktitle}{\emph{Proceedings of the 40th International Conference on Machine Learning}} (Honolulu, Hawaii, USA). \bibinfo{pages}{19730--19742}.
\newblock


\bibitem[Li et~al\mbox{.}(2022)]%
        {li2022blip}
\bibfield{author}{\bibinfo{person}{Junnan Li}, \bibinfo{person}{Dongxu Li}, \bibinfo{person}{Caiming Xiong}, {and} \bibinfo{person}{Steven Hoi}.} \bibinfo{year}{2022}\natexlab{}.
\newblock \bibinfo{title}{BLIP: Bootstrapping Language-Image Pre-training for Unified Vision-Language Understanding and Generation}.
\newblock
\showeprint[arxiv]{2201.12086}~[cs.CV]


\bibitem[Li et~al\mbox{.}(2021)]%
        {li2021albef}
\bibfield{author}{\bibinfo{person}{Junnan Li}, \bibinfo{person}{Ramprasaath~R. Selvaraju}, \bibinfo{person}{Akhilesh~D. Gotmare}, \bibinfo{person}{Shafiq Joty}, \bibinfo{person}{Caiming Xiong}, {and} \bibinfo{person}{Steven~C.H. Hoi}.} \bibinfo{year}{2021}\natexlab{}.
\newblock \showarticletitle{Align before fuse: vision and language representation learning with momentum distillation}. In \bibinfo{booktitle}{\emph{Proceedings of the 35th International Conference on Neural Information Processing Systems}}. \bibinfo{address}{Red Hook, NY, USA}.
\newblock


\bibitem[Liang et~al\mbox{.}(2024)]%
        {liang2024poir}
\bibfield{author}{\bibinfo{person}{Zihan Liang}, \bibinfo{person}{Ben Chen}, \bibinfo{person}{Zhuoran Ran}, \bibinfo{person}{Zihan Wang}, \bibinfo{person}{Huangyu Dai}, \bibinfo{person}{Yufei Ma}, \bibinfo{person}{Dehong Gao}, \bibinfo{person}{Xiaoyan Cai}, {and} \bibinfo{person}{Libin Yang}.} \bibinfo{year}{2024}\natexlab{}.
\newblock \showarticletitle{Self-Renewal Prompt Optimizing with Implicit Reasoning}. In \bibinfo{booktitle}{\emph{Findings of the Association for Computational Linguistics: EMNLP 2024}}. \bibinfo{address}{Miami, Florida, USA}, \bibinfo{pages}{3030--3041}.
\newblock


\bibitem[Lin et~al\mbox{.}(2025)]%
        {lin2025mmembed}
\bibfield{author}{\bibinfo{person}{Sheng-Chieh Lin}, \bibinfo{person}{Chankyu Lee}, \bibinfo{person}{Mohammad Shoeybi}, \bibinfo{person}{Jimmy Lin}, \bibinfo{person}{Bryan Catanzaro}, {and} \bibinfo{person}{Wei Ping}.} \bibinfo{year}{2025}\natexlab{}.
\newblock \showarticletitle{{MM}-{EMBED}: {UNIVERSAL} {MULTIMODAL} {RETRIEVAL} {WITH} {MULTIMODAL} {LLMS}}. In \bibinfo{booktitle}{\emph{The Thirteenth International Conference on Learning Representations}}.
\newblock


\bibitem[Liu et~al\mbox{.}(2023)]%
        {liu2023univldr}
\bibfield{author}{\bibinfo{person}{Zhenghao Liu}, \bibinfo{person}{Chenyan Xiong}, \bibinfo{person}{Yuanhuiyi Lv}, \bibinfo{person}{Zhiyuan Liu}, {and} \bibinfo{person}{Ge Yu}.} \bibinfo{year}{2023}\natexlab{}.
\newblock \showarticletitle{Universal Vision-Language Dense Retrieval: Learning A Unified Representation Space for Multi-Modal Retrieval}. In \bibinfo{booktitle}{\emph{The Eleventh International Conference on Learning Representations}}.
\newblock


\bibitem[Mao et~al\mbox{.}(2023)]%
        {Mao2023CELoss}
\bibfield{author}{\bibinfo{person}{Anqi Mao}, \bibinfo{person}{Mehryar Mohri}, {and} \bibinfo{person}{Yutao Zhong}.} \bibinfo{year}{2023}\natexlab{}.
\newblock \showarticletitle{Cross-entropy loss functions: theoretical analysis and applications}. In \bibinfo{booktitle}{\emph{Proceedings of the 40th International Conference on Machine Learning}} (Honolulu, Hawaii, USA).
\newblock


\bibitem[Radford et~al\mbox{.}(2021)]%
        {radford2021clip}
\bibfield{author}{\bibinfo{person}{Alec Radford}, \bibinfo{person}{Jong~Wook Kim}, \bibinfo{person}{Chris Hallacy}, \bibinfo{person}{Aditya Ramesh}, \bibinfo{person}{Gabriel Goh}, \bibinfo{person}{Sandhini Agarwal}, \bibinfo{person}{Girish Sastry}, \bibinfo{person}{Amanda Askell}, \bibinfo{person}{Pamela Mishkin}, \bibinfo{person}{Jack Clark}, \bibinfo{person}{Gretchen Krueger}, {and} \bibinfo{person}{Ilya Sutskever}.} \bibinfo{year}{2021}\natexlab{}.
\newblock \bibinfo{title}{Learning Transferable Visual Models From Natural Language Supervision}.
\newblock
\showeprint[arxiv]{2103.00020}~[cs.CV]


\bibitem[Schroff et~al\mbox{.}(2015)]%
        {Florian2015tripletloss}
\bibfield{author}{\bibinfo{person}{Florian Schroff}, \bibinfo{person}{Dmitry Kalenichenko}, {and} \bibinfo{person}{James Philbin}.} \bibinfo{year}{2015}\natexlab{}.
\newblock \showarticletitle{FaceNet: A unified embedding for face recognition and clustering}. In \bibinfo{booktitle}{\emph{2015 IEEE Conference on Computer Vision and Pattern Recognition (CVPR)}}. \bibinfo{pages}{815--823}.
\newblock


\bibitem[Wang et~al\mbox{.}(2024)]%
        {wang2024qwen2vl}
\bibfield{author}{\bibinfo{person}{Peng Wang}, \bibinfo{person}{Shuai Bai}, \bibinfo{person}{Sinan Tan}, \bibinfo{person}{Shijie Wang}, \bibinfo{person}{Zhihao Fan}, \bibinfo{person}{Jinze Bai}, \bibinfo{person}{Keqin Chen}, \bibinfo{person}{Xuejing Liu}, \bibinfo{person}{Jialin Wang}, \bibinfo{person}{Wenbin Ge}, \bibinfo{person}{Yang Fan}, \bibinfo{person}{Kai Dang}, \bibinfo{person}{Mengfei Du}, \bibinfo{person}{Xuancheng Ren}, \bibinfo{person}{Rui Men}, \bibinfo{person}{Dayiheng Liu}, \bibinfo{person}{Chang Zhou}, \bibinfo{person}{Jingren Zhou}, {and} \bibinfo{person}{Junyang Lin}.} \bibinfo{year}{2024}\natexlab{}.
\newblock \bibinfo{title}{Qwen2-VL: Enhancing Vision-Language Model's Perception of the World at Any Resolution}.
\newblock
\showeprint[arxiv]{2409.12191}~[cs.CV]


\bibitem[Wei et~al\mbox{.}(2024)]%
        {Wei2024uniir}
\bibfield{author}{\bibinfo{person}{Cong Wei}, \bibinfo{person}{Yang Chen}, \bibinfo{person}{Haonan Chen}, \bibinfo{person}{Hexiang Hu}, \bibinfo{person}{Ge Zhang}, \bibinfo{person}{Jie Fu}, \bibinfo{person}{Alan Ritter}, {and} \bibinfo{person}{Wenhu Chen}.} \bibinfo{year}{2024}\natexlab{}.
\newblock \showarticletitle{UniIR: Training and Benchmarking Universal Multimodal Information Retrievers}. In \bibinfo{booktitle}{\emph{Computer Vision – ECCV 2024: 18th European Conference}} (Milan, Italy). \bibinfo{pages}{387--404}.
\newblock


\bibitem[Wu et~al\mbox{.}(2020)]%
        {wu2020fashioniqnewdataset}
\bibfield{author}{\bibinfo{person}{Hui Wu}, \bibinfo{person}{Yupeng Gao}, \bibinfo{person}{Xiaoxiao Guo}, \bibinfo{person}{Ziad Al-Halah}, \bibinfo{person}{Steven Rennie}, \bibinfo{person}{Kristen Grauman}, {and} \bibinfo{person}{Rogerio Feris}.} \bibinfo{year}{2020}\natexlab{}.
\newblock \bibinfo{title}{Fashion IQ: A New Dataset Towards Retrieving Images by Natural Language Feedback}.
\newblock
\showeprint[arxiv]{1905.12794}~[cs.CV]


\bibitem[Wu et~al\mbox{.}(2024)]%
        {wu2024deepseekvl2}
\bibfield{author}{\bibinfo{person}{Zhiyu Wu}, \bibinfo{person}{Xiaokang Chen}, \bibinfo{person}{Zizheng Pan}, \bibinfo{person}{Xingchao Liu}, \bibinfo{person}{Wen Liu}, \bibinfo{person}{Damai Dai}, \bibinfo{person}{Huazuo Gao}, \bibinfo{person}{Yiyang Ma}, \bibinfo{person}{Chengyue Wu}, \bibinfo{person}{Bingxuan Wang}, \bibinfo{person}{Zhenda Xie}, \bibinfo{person}{Yu Wu}, \bibinfo{person}{Kai Hu}, \bibinfo{person}{Jiawei Wang}, \bibinfo{person}{Yaofeng Sun}, \bibinfo{person}{Yukun Li}, \bibinfo{person}{Yishi Piao}, \bibinfo{person}{Kang Guan}, \bibinfo{person}{Aixin Liu}, \bibinfo{person}{Xin Xie}, \bibinfo{person}{Yuxiang You}, \bibinfo{person}{Kai Dong}, \bibinfo{person}{Xingkai Yu}, \bibinfo{person}{Haowei Zhang}, \bibinfo{person}{Liang Zhao}, \bibinfo{person}{Yisong Wang}, {and} \bibinfo{person}{Chong Ruan}.} \bibinfo{year}{2024}\natexlab{}.
\newblock \bibinfo{title}{DeepSeek-VL2: Mixture-of-Experts Vision-Language Models for Advanced Multimodal Understanding}.
\newblock
\showeprint[arxiv]{2412.10302}~[cs.CV]


\bibitem[Yang et~al\mbox{.}(2023)]%
        {yang2023gpt4v}
\bibfield{author}{\bibinfo{person}{Zhengyuan Yang}, \bibinfo{person}{Linjie Li}, \bibinfo{person}{Kevin Lin}, \bibinfo{person}{Jianfeng Wang}, \bibinfo{person}{Chung-Ching Lin}, \bibinfo{person}{Zicheng Liu}, {and} \bibinfo{person}{Lijuan Wang}.} \bibinfo{year}{2023}\natexlab{}.
\newblock \bibinfo{title}{The Dawn of LMMs: Preliminary Explorations with GPT-4V(ision)}.
\newblock
\showeprint[arxiv]{2309.17421}~[cs.CV]


\bibitem[Yao et~al\mbox{.}(2023)]%
        {yao2023react}
\bibfield{author}{\bibinfo{person}{Shunyu Yao}, \bibinfo{person}{Jeffrey Zhao}, \bibinfo{person}{Dian Yu}, \bibinfo{person}{Nan Du}, \bibinfo{person}{Izhak Shafran}, \bibinfo{person}{Karthik~R Narasimhan}, {and} \bibinfo{person}{Yuan Cao}.} \bibinfo{year}{2023}\natexlab{}.
\newblock \showarticletitle{ReAct: Synergizing Reasoning and Acting in Language Models}. In \bibinfo{booktitle}{\emph{The Eleventh International Conference on Learning Representations}}.
\newblock


\bibitem[Zhang et~al\mbox{.}(2024)]%
        {zhang2024gme}
\bibfield{author}{\bibinfo{person}{Xin Zhang}, \bibinfo{person}{Yanzhao Zhang}, \bibinfo{person}{Wen Xie}, \bibinfo{person}{Mingxin Li}, \bibinfo{person}{Ziqi Dai}, \bibinfo{person}{Dingkun Long}, \bibinfo{person}{Pengjun Xie}, \bibinfo{person}{Meishan Zhang}, \bibinfo{person}{Wenjie Li}, {and} \bibinfo{person}{Min Zhang}.} \bibinfo{year}{2024}\natexlab{}.
\newblock \bibinfo{title}{GME: Improving Universal Multimodal Retrieval by Multimodal LLMs}.
\newblock
\showeprint[arxiv]{2412.16855}~[cs.CL]


\bibitem[Zhao et~al\mbox{.}(2024)]%
        {Zhao2024ExpelAgent}
\bibfield{author}{\bibinfo{person}{Andrew Zhao}, \bibinfo{person}{Daniel Huang}, \bibinfo{person}{Quentin Xu}, \bibinfo{person}{Matthieu Lin}, \bibinfo{person}{Yong-Jin Liu}, {and} \bibinfo{person}{Gao Huang}.} \bibinfo{year}{2024}\natexlab{}.
\newblock \showarticletitle{ExpeL: LLM Agents Are Experiential Learners}. In \bibinfo{booktitle}{\emph{Proceedings of the AAAI Conference on Artificial Intelligence}}. \bibinfo{pages}{19632--19642}.
\newblock


\bibitem[Zhou et~al\mbox{.}(2024a)]%
        {zhou2024bgevl}
\bibfield{author}{\bibinfo{person}{Junjie Zhou}, \bibinfo{person}{Zheng Liu}, \bibinfo{person}{Ze Liu}, \bibinfo{person}{Shitao Xiao}, \bibinfo{person}{Yueze Wang}, \bibinfo{person}{Bo Zhao}, \bibinfo{person}{Chen~Jason Zhang}, \bibinfo{person}{Defu Lian}, {and} \bibinfo{person}{Yongping Xiong}.} \bibinfo{year}{2024}\natexlab{a}.
\newblock \bibinfo{title}{MegaPairs: Massive Data Synthesis For Universal Multimodal Retrieval}.
\newblock
\showeprint[arxiv]{2412.14475}~[cs.CV]


\bibitem[Zhou et~al\mbox{.}(2024b)]%
        {zhou-etal-2024-vista}
\bibfield{author}{\bibinfo{person}{Junjie Zhou}, \bibinfo{person}{Zheng Liu}, \bibinfo{person}{Shitao Xiao}, \bibinfo{person}{Bo Zhao}, {and} \bibinfo{person}{Yongping Xiong}.} \bibinfo{year}{2024}\natexlab{b}.
\newblock \showarticletitle{{VISTA}: Visualized Text Embedding For Universal Multi-Modal Retrieval}. In \bibinfo{booktitle}{\emph{Proceedings of the 62nd Annual Meeting of the Association for Computational Linguistics (Volume 1: Long Papers)}} (Bangkok, Thailand). \bibinfo{pages}{3185--3200}.
\newblock


\bibitem[Zhou et~al\mbox{.}(2024c)]%
        {zhou-etal-2024-marvel}
\bibfield{author}{\bibinfo{person}{Tianshuo Zhou}, \bibinfo{person}{Sen Mei}, \bibinfo{person}{Xinze Li}, \bibinfo{person}{Zhenghao Liu}, \bibinfo{person}{Chenyan Xiong}, \bibinfo{person}{Zhiyuan Liu}, \bibinfo{person}{Yu Gu}, {and} \bibinfo{person}{Ge Yu}.} \bibinfo{year}{2024}\natexlab{c}.
\newblock \showarticletitle{{MARVEL}: Unlocking the Multi-Modal Capability of Dense Retrieval via Visual Module Plugin}. In \bibinfo{booktitle}{\emph{Proceedings of the 62nd Annual Meeting of the Association for Computational Linguistics (Volume 1: Long Papers)}} (Bangkok, Thailand). \bibinfo{pages}{14608--14624}.
\newblock


\bibitem[Zhu et~al\mbox{.}(2024)]%
        {Zhu2024MIMAmazonKDD24}
\bibfield{author}{\bibinfo{person}{Xinliang Zhu}, \bibinfo{person}{Sheng~Wei Huang}, \bibinfo{person}{Han Ding}, \bibinfo{person}{Jinyu Yang}, \bibinfo{person}{Kelvin Chen}, \bibinfo{person}{Tao Zhou}, \bibinfo{person}{Tal Neiman}, \bibinfo{person}{Ouye Xie}, \bibinfo{person}{Son Tran}, \bibinfo{person}{Benjamin Yao}, \bibinfo{person}{Douglas Gray}, \bibinfo{person}{Anuj Bindal}, {and} \bibinfo{person}{Arnab Dhua}.} \bibinfo{year}{2024}\natexlab{}.
\newblock \showarticletitle{Bringing Multimodality to Amazon Visual Search System}. In \bibinfo{booktitle}{\emph{Proceedings of the 30th ACM SIGKDD Conference on Knowledge Discovery and Data Mining}} (Barcelona, Spain). \bibinfo{pages}{6390--6399}.
\newblock


\end{thebibliography}

\appendix

\end{document}